\documentclass[a4paper,twocolumn,longbibliography, aps,pra,superscriptaddress,amsfonts,amsmath,amssymb,floatfix]{revtex4-2}

\usepackage[
  colorlinks,
  citecolor=blue,
  linkcolor=blue,
  urlcolor=blue
]{hyperref}
\usepackage{graphicx,epstopdf, bm}
\usepackage{braket}
\usepackage{blindtext}
\usepackage{enumitem}
\usepackage[capitalize]{cleveref}
\usepackage{xcolor}
\usepackage[all]{hypcap}
\usepackage{cleveref}
\usepackage{amssymb}
\crefname{section}{Sec.}{Secs.}
\Crefname{section}{Section}{Sections}
\crefname{appendix}{App.}{Appces.}
\Crefname{appendix}{Appendix}{Appendices}

\begin{document}

\title{The hidden negative differential thermal conductance
}


\author{Zi-chen Zhang, Zheng Liu, and Chang-shui Yu}
	\email{Electronic address: ycs@dlut.edu.cn}
	\address{School of Physics, Dalian University of Technology, Dalian 116024, P.R. China}
\date{\today}

	\begin{abstract}
Negative differential thermal conductance (NDTC), a hallmark of nonlinear quantum thermal transport, plays a critical role in the design of quantum thermal devices such as thermal diodes and transistors. The Lindblad dynamics predicts that the heat current through two coupled atoms increases with the increasing temperature difference of two bosonic reservoirs. However, in this paper, we uncover the suppressive effect on the heat current in this nonequilibrium system using the Bloch-Redfield master equations, which indicate the emergence of NDTC. Our findings underscore the crucial role of beyond-Lindblad dynamics in accurately capturing nonlinear features in quantum thermodynamic systems.
	\end{abstract}


\maketitle

\section{Introduction}

Quantum thermodynamics provides the foundational framework for understanding energy conversion processes at the microscopic scale, where quantum coherence, entanglement, or other quantum phenomena may play a significant role. This theoretical framework, elucidating the interconversion of work and heat in quantum systems, forms the basis for quantum thermal devices, such as quantum engines \cite{PhysRevE.61.4774, PhysRevE.73.026109,doi:10.1063/1.461951, PhysRevLett.125.166802, PhysRevLett.124.110604,PhysRevB.101.081408,e22111255, PhysRevLett.122.150604,PhysRevE.106.014143, Colla_2025}, quantum refrigerators \cite{PhysRevE.64.056130,doi:10.1119/1.18197, Skrzypczyk_2011,yu2014re,yu2019quantum, PhysRevLett.122.070603,PhysRevLett.129.030601}, and work extraction using quantum measurements \cite{PhysRevResearch.4.033103, PhysRevLett.118.260603, PhysRevA.106.042221, PhysRevA.80.012322, PhysRevA.106.052201}, which demonstrates how quantum resources can be harnessed to control energy flows at the quantum level. Essentially, describing such systems requires a careful treatment of heat currents in open quantum systems, especially when they are driven far from equilibrium by temperature gradients between reservoirs. The Lindblad master equation, based on the Born-Markovian approximation and the secular approximation (ME), has been widely adopted for its simplicity and complete positivity in its ability to describe the dynamics of open quantum systems effectively \cite{breuer2002theory,weiss2012quantum,carmichael2013statistical,schaller2014non,lidar2019lecture, PhysRev.138.B1007,rivas2010markovian,albash2012quantum}, especially including quantum optics \cite{gardiner2004quantum,mozgunov2020completely}, nuclear magnetic resonance \cite{harris1996encyclopedia}, quantum transport \cite{ishizaki2009adequacy} and chemical dynamical systems \cite{pollard1996redfield} attesting to the versatility in describing system-environment interactions \cite{gemmer2009quantum,alicki2006internal, PhysRevLett.122.150603, PhysRevE.107.014108}.

The Lindblad ME is typically valid for weak system-bath coupling and large energy separations. Still, it fails in systems with small energy gaps, where coherence-mediated processes and non-secular energy exchange become significant. In contrast, the Bloch-Redfield (BR) master equation \cite{REDFIELD19651,redfield1957theory,lindblad1976generators,davies1974markovian,gorini1976completely,PhysRevB.103.214308,mozgunov2020completely,davidovic2020completely,PhysRevE.106.054145} derived from second-order perturbation theory provides a more complete and accurate description of dissipative processes by retaining non-secular terms \cite{redfield1957theory,lindblad1976generators,davies1974markovian}. In particular, the BR master equation's steady-state solution corresponds to the mean force Gibbs (MFG) state \cite{10.1116/5.0073853, PhysRevLett.127.250601,10.1063/1.4718706, PhysRevE.106.054145}, which properly incorporates system-environment coupling effects through a consistent perturbative expansion. The MFG state of the BR master equation contains the second-order corrections in the system-environment coupling that modify both populations and coherences. In contrast, the Lindblad solution retains only the zeroth-order Gibbs distribution.

In this paper, we study the negative differential thermal conductance (NDTC)—a counterintuitive phenomenon in which increasing the temperature difference between thermal reservoirs leads to a decrease in the steady state heat current \cite{PhysRevB.73.205415, PhysRevB.103.075407, PhysRevB.107.134510, PhysRevE.99.032114}. NDTC is not only of fundamental interest in nonequilibrium quantum thermodynamics but also a key enabler for functional microscopic thermal devices, such as quantum thermal diodes and transistors \cite{joulain2015modulation, RevModPhys.84.1045, PhysRevE.95.022128, PhysRevE.99.042121, PhysRevE.99.042102, Tesser_2022, PhysRevLett.120.200603}. Our system is composed of two coupled atoms interacting with nonequilibrium thermal reservoirs \cite{PhysRevLett.100.105901,PhysRevLett.88.094302,PhysRevE.83.031106,PhysRevB.100.045418,hovhannisyan2019quantum,PhysRevA.75.032308,PhysRevA.78.062301,yang2022heat,hu2018steady,cattaneo2019local,PhysRevA.99.042320}.  Through perturbative analysis, we derive the BR master equation with the zeroth-order term recovering the Lindblad steady state, while second-order corrections reflect equilibrium features consistent with the MFG state. We uncover a suppressive effect on heat current arising from non-secular terms, which gives rise to NDTC in specific parameter regimes, even though the Lindblad dynamics predicts an increasing heat current with the temperature difference. The Redfield approach proves particularly powerful for describing nonequilibrium steady states and heat transport, providing crucial insights into the bridge between physically accurate dynamics and Markovian approximations for optimized quantum thermal control. This paper is organized as follows. In Sec. \ref{2}, we give a brief description of our model and derive the master equations under the Born-Markov approximation. In Sec. \ref{3}, we focus on calculating the steady states. In Sec. \ref{4}, we compare the heat currents derived from Lindblad and BR master equations. We conclude with a summary In Sec. \ref{5}.


\section{The Nonequilibrium open system}
\label{2}
\begin{figure}[t]
\centering\includegraphics[width=9cm]{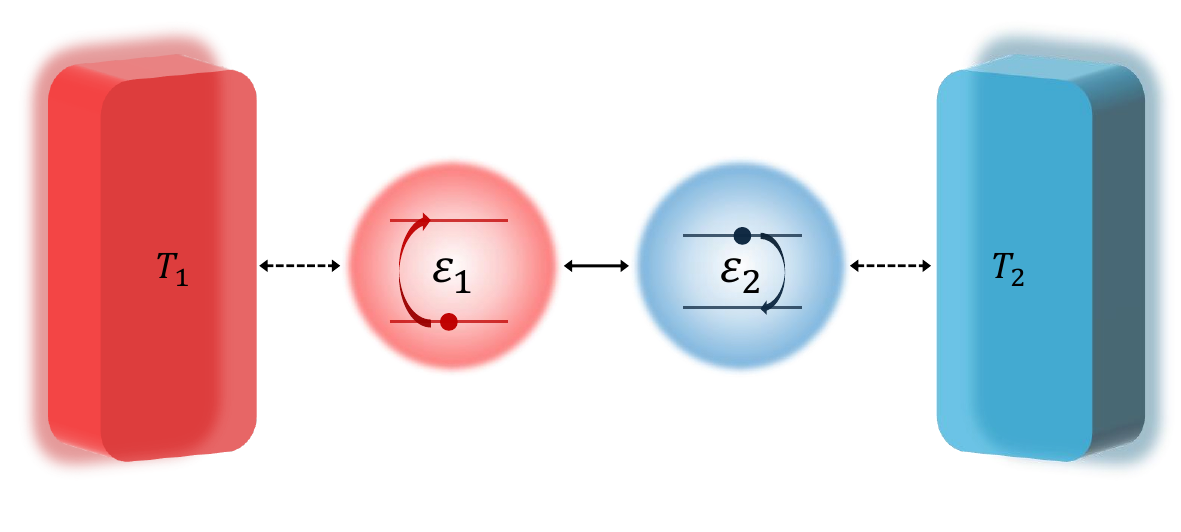}
\caption{The schematic illustration of our model, where the dashed line
represents weak coupling, and the solid line represents strong coupling. The
temperatures of two heat reservoirs are $T_{1}$ and $T_{2}$, the energy
separation of two qubits are $\protect\varepsilon _{1}$ and $\protect%
\varepsilon _{2}$.}
\label{example}
\end{figure}

Our system consists of two coupled two-level atoms (TLAs) interacting with a
distinct thermal reservoir, respectively, as shown in Fig. \ref{example}.
The
Hamiltonian of the TLAs system reads
\begin{equation}
{H_{S}}=\frac{{\varepsilon _{1}}}{2}\sigma _{1}^{z}+\frac{{\varepsilon _{2}}%
}{2}\sigma _{2}^{z}+g \sigma _{1}^{x}\sigma _{2}^{x},
\end{equation}
where $\sigma _{i}^{z}$ and $\sigma _{i}^{x}$ are the Pauli matrices, and $g 
$ is the coupling strength of two qubits, and we have set the Planck constant $\hbar =1$
and the Boltzmann constant $k_{B}=1$. Without loss of generality, let $%
\varepsilon _{1}\geq \varepsilon _{2}$, then we have the eigenvalues of ${H%
}_{S}$ as ${s_{1}}=\beta ,{s_{4}}=-\beta ,{s_{2}}=\alpha ,{s_{3}}=-\alpha ,$
where 
\begin{equation}
\alpha =\sqrt{\frac{{{{\left( {{\varepsilon _{1}}-{\varepsilon _{2}}}\right) 
}^{2}}}}{4}+{g ^{2}}},\beta =\sqrt{\frac{{{{\left( {{\varepsilon _{1}}+{%
\varepsilon _{2}}}\right) }^{2}}}}{4}+{g ^{2}}}.
\end{equation}
The corresponding eigenstates are 
\[\begin{gathered}
\left| {{s_1}} \right\rangle  = \sin \frac{\varphi }{2}\left| {0,0} \right\rangle  + \cos \frac{\varphi }{2}\left| {1,1} \right\rangle , \hfill\\
\left| {{s_4}} \right\rangle  =\cos \frac{\varphi }{2}\left| {0,0} \right\rangle  - \sin \frac{\varphi }{2}\left| {1,1} \right\rangle, \hfill \\
  \left| {{s_2}} \right\rangle  = \cos \frac{\theta }{2}\left| {1,0} \right\rangle  + \sin \frac{\theta }{2}\left| {0,1} \right\rangle ,\hfill\\
  \left| {{s_3}} \right\rangle  =  - \sin \frac{\theta }{2}\left| {1,0} \right\rangle  + \cos \frac{\theta }{2}\left| {0,1} \right\rangle , \hfill \\ 
\end{gathered} \]
with 
\begin{equation}
\tan \varphi =\frac{{2g }}{{{\varepsilon _{1}}+{\varepsilon _{2}}}},\tan
\theta =\frac{{2g }}{{{\varepsilon _{1}}-{\varepsilon _{2}}}}.
\end{equation}
Suppose that the two heat reservoirs are always in the thermal state with
inverse temperature $\beta _{j}=1/T_{j}$, then the Hamiltonian of the
reservoirs are 
\begin{equation}
{H}_{Bj}=\sum_{n}\omega _{n,j}{b}_{j,n}^{\dagger }{b}_{j,n},
\end{equation}%
where the summation is for all the harmonic oscillator modes of the $j$th reservoir. The interaction between the system and the reservoir is given by 
\begin{align}
{H_{S{B_{j}}}}=\lambda_j B_{j}^{x} \sigma _{j}^{x} = \lambda_j B_{j}^{x}\sum\limits_{\mu }{ {{V_j}\left( {{\omega _\mu }} \right)} ,}
\end{align}
where $\lambda_j$ is a dimensionless coupling constant that characterizes the strength of the interaction between the j-th atom and the j-th reservoir, $B_{j}^{x}=\sum\limits_{n}{{\kappa_{j,n}} ( {{b_{j,n}}+b_{j,n}^{\dag }} ) }$ and $\sigma _j^x = \sum\limits_{{\mu }} {{V_j}\left( {{\omega _\mu }} \right)} $. Here ${{V_j}\left( {{\omega _\mu }} \right)}$ denote the eigenoperators satisfying $\left[ {{H_S},{V_j}\left( {{\omega _\mu }} \right)} \right] =  - {\omega _\mu }{V_j}\left( {{\omega _\mu }} \right)$ \cite{breuer2002theory,schaller2014non}, given as 
\[\begin{gathered}
  {V_1}\left( {{\omega _1}} \right) = \sin {\phi _ + }\left( {\left| {{s_2}} \right\rangle \left\langle {{s_1}} \right| - \left| {{s_4}} \right\rangle \left\langle {{s_3}} \right|} \right),\hfill\\
  {V_1}\left( {{\omega _2}} \right) = \cos {\phi _ - }\left( {\left| {{s_4}} \right\rangle \left\langle {{s_2}} \right| + \left| {{s_3}} \right\rangle \left\langle {{s_1}} \right|} \right), \hfill \\
  {V_2}\left( {{\omega _1}} \right) = \cos {\phi _ - }\left( {\left| {{s_2}} \right\rangle \left\langle {{s_1}} \right| + \left| {{s_4}} \right\rangle \left\langle {{s_3}} \right|} \right),\hfill\\
  {V_2}\left( {{\omega _2}} \right) = \sin {\phi _ + }\left( {\left| {{s_4}} \right\rangle \left\langle {{s_2}} \right| - \left| {{s_3}} \right\rangle \left\langle {{s_1}} \right|} \right), \hfill \\ 
\end{gathered} \]
with 
\begin{equation}
{\phi _{+}}=\frac{{\theta +\varphi }}{2},{\phi _{-}}=\frac{{\theta -\varphi }%
}{2},\omega_1=\beta-\alpha,\omega_2=\beta+\alpha,
\end{equation}
and the index number ${ \mu } \in \left\{ 1,-1,2,-2 \right\}$, noting that the negative frequency operator is given by ${V_j}\left( {{\omega _{-\mu} }} \right) = {V_j}\left( { - {\omega _\mu }} \right) = V_j^\dag \left( {{\omega _\mu }} \right)$.

Thus, we can write the total Hamiltonian of the system and the reservoirs as 
\begin{equation}
{H}={H}_{S}+{H}_{B1}+{H}_{B2}+{H}_{SB1}+{H}_{SB2}.
\end{equation}
 Following the standard process \cite{breuer2002theory,schaller2014non} in the framework of Born-Markov approximation, one can directly get the BR master equation governing the evolution of the system as 
\begin{equation}
\frac{d{\rho }}{dt}=-i[{H}_{S}+H_{LS},{\rho }]+\mathcal{L}_{1}({\rho })+%
\mathcal{L}_{2}({\rho }),  \label{master}
\end{equation}%
where $\rho $ is the density matrix of the TLAs, $H_{LS}$ is the energy correction,
i.e., the Lamb shift
\begin{equation}
{H_{LS}} = \sum\limits_j { \sum\limits_{\omega ,\omega '} {{\xi _j}\left( {\omega ,\omega '} \right)V_j^\dag \left( {\omega '} \right){V_j}\left( \omega  \right)} } ,
\end{equation}
and $\mathcal{L}_{j}({\rho })$ are the dissipators given by 
\begin{eqnarray}
&&\mathcal{L}_j\left( \rho  \right) = \sum\limits_{\omega ,\omega '} {\gamma _j}\left( {\omega ,\omega '} \right)\notag\\
&&\times\left( {V_j}\left( {{\omega }} \right)\rho V_j^\dag \left( {\omega '} \right) - \frac{1}{2}\left\{ {V_j^\dag \left( {\omega '} \right){V_j}\left( \omega  \right),\rho } \right\} \right).
\end{eqnarray}
The coefficients are defined as
\begin{eqnarray}
{\xi _j}\left( {\omega ,\omega '} \right) =\lambda_j^2 \frac{{{\Gamma _j}\left( \omega  \right) - \Gamma _j^*\left( {\omega '} \right)}}{{2i}},\\{\gamma _j}\left( {\omega ,\omega '} \right) = \lambda_j^2 \left({\Gamma _j}\left( \omega  \right) + \Gamma _j^*\left( {\omega '} \right)\right),
\label{xi}
\end{eqnarray}
where ${\Gamma _{j}}\left( \omega \right) =\int_{0}^{\infty }{ds%
{e^{i\omega s}}\left\langle {B_{j}^{x}\left( s\right) B_{j}^{x}}%
\right\rangle }$ is the forward Fourier transform of the reservoir correlation
function $\left\langle {B_{j}^{x}\left( s\right) B_{j}^{x}}\right\rangle $. One can write 
\begin{gather}
  {\Gamma _j}\left( \omega  \right) \equiv {G_j}\left( \omega  \right) + i{S_j}\left( \omega  \right), \hfill \\
  {G_j}\left( \omega  \right) = {J_j}\left( \omega  \right)\left( {{{\overline n }_j}\left( \omega  \right) + 1} \right), \hfill \\
  {S_j}\left( \omega  \right) = \frac{1}{\pi }P.V.\int_0^\infty  {{J_j}\left( {\omega '} \right)\left( {\frac{{{{\overline n }_j}\left( {\omega '} \right) + 1}}{{\omega  - \omega '}} + \frac{{{{\overline n }_j}\left( {\omega '} \right)}}{{\omega  + \omega '}}} \right)d\omega '} ,
\end{gather}
where the $P.V.$ denotes the Cauchy principal value of the integral, $\bar{n}_{j}(\omega )={(\exp(\beta _{j}\omega )-1)}^{-1}$ are the average photon number,  and
\begin{equation}
J_{j}\left( \omega \right) =\pi {\sum\limits_{n}{\left\vert {\kappa_{j,n}}\right\vert }^{2}}\delta \left( {\omega -{\omega _{n}}}\right)
\end{equation}
are the spectral density of the heat reservoirs. A more
detailed derivation is provided in the Appendix \ref{appendixA}. Usually, we treat the
superposition of these infinitely many delta functions as a continuous
function, for example, the spectral density of an Ohmic-type heat reservoir
with a high cut-off frequency $\omega _{D}$ considered in this paper takes
the form \cite{breuer2002theory,weiss2012quantum} 
\begin{equation}
{J_{j}}\left( \omega \right) =\frac{{{C_{j}}\omega }}{{1+{{\left( {%
\omega /{\omega _{D}}}\right) }^{2}}}}.  \label{density}
\end{equation}%
This canonical Drude cut-off is an extension of the Ohmic model, where the
spectral density is modified to include a frequency-dependent damping
factor. This damping factor is often represented as a Lorentzian function,
introduces a high-frequency cut-off, effectively limiting the spectral
density beyond a certain frequency. This modification is crucial for making
the model more realistic by preventing the unphysical divergence of the
spectral density at high frequencies. Recalling Eq. (\ref{xi}) and noting that $\lambda_j^2$ and $J_j(\omega)$ always appear together as a product in the master equation, we absorb $\lambda_j^2$ into $J_j(\omega)$ for simplicity by setting $\gamma_j \equiv C_j \lambda_j^2$ and replacing $C_j$ in Eq. (\ref{density}) with $\gamma_j$.

\section{The steady states}
\label{3}
Based on the master equation (\ref{master}), one can obtain the steady state by solving $\frac{{d\rho }}{{dt}} = 0$.  As derived in Appendix \ref{appendixB}, the steady state of the system in the eigenbasis of $H_S$ has the form of 
\begin{equation}
{\rho _S} = \left( {\begin{array}{*{20}{c}}
  {{\rho _{11}}}&{}&{}&{{\rho _{14}}} \\ 
  {}&{{\rho _{22}}}&{{\rho _{23}}}&{} \\ 
  {}&{{\rho _{32}}}&{{\rho _{33}}}&{} \\ 
  {{\rho _{41}}}&{}&{}&{{\rho _{44}}} 
\end{array}} \right).
\end{equation}
Instead of numerically calculating $\rho _S$, we analyzed the steady state using a perturbative approach in Appendix \ref{appendixB} since the explicit analytical expression is difficult to give directly. We find that $\rho _S^{(0)}$ coincides precisely with the steady state derived from the Lindblad master equation (\ref{A14}) under the same parameter regime, while $\rho _S^{(2)}$ provides the second-order correction. Regarding positively defined $\rho_S$, since the steady state $\rho _S^{(0)}$ corresponding to the Lindblad master equation is positive, second-order corrections must preserve the positivity of $\rho_S$, which is guaranteed by the weak coupling condition $\gamma_j \ll 1$.

\begin{figure}[t]
\centering\includegraphics[width=9cm]{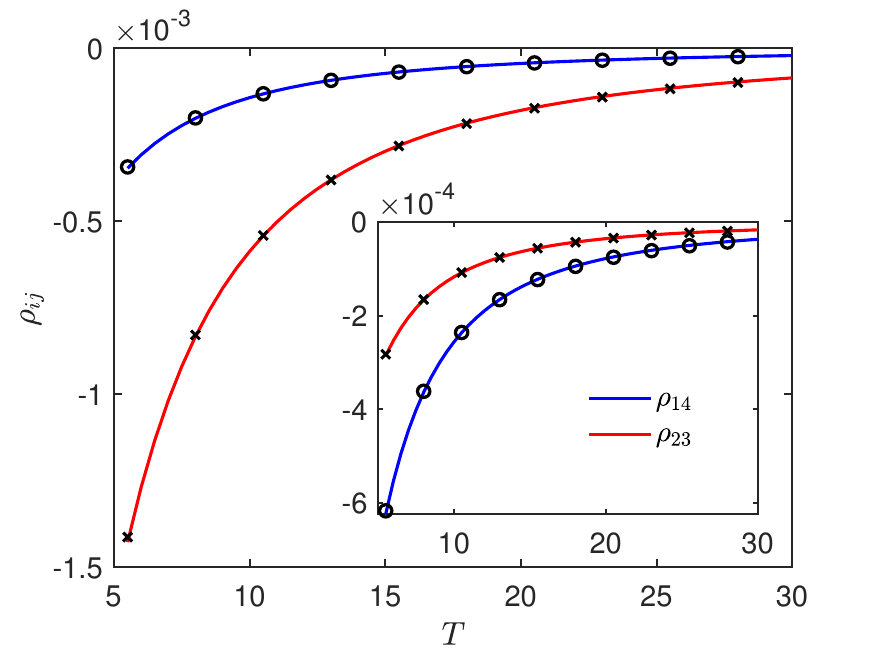}
\caption{The secondary diagonal elements of the system's steady state or the mean force Gibbs state $\rho_{ij}$ dependent on the temperature of the thermal reservoirs $T_1=T_2=T$. Here we set $\gamma _{1}=0.01,\gamma _{2}=0,\omega _{D}=50,\varepsilon _{1}=5,\varepsilon _{2}=4,g=1$, only the first thermal reservoir interacts with the system. The blue line represents $\rho_{14}$; the red line represents $\rho_{23}$, and the 'o and 'x' labeled points represent the corresponding $\rho_{ij}$ of the mean force Gibbs state. For the inset, $\gamma _{2}=0.01$, while keeping all other parameters unchanged, both thermal reservoirs interact with the system for this case.}
\label{fig3}
\end{figure}

To verify the master equation, we first consider two special cases. For $\gamma_2=0$ corresponding to the case where the system is coupled to only one of the two thermal reservoirs,  the usual expectation is that the steady state of the system is the Gibbs state, which corresponds to the Lindblad steady state as
\begin{equation}
\rho _S^{\left( 0 \right)} = {\tau _S} \equiv \frac{{\exp \left( { - {\beta_1}{H_S}} \right)}}{{Tr\left( {\exp \left( { - {\beta_1}{H_S}} \right)} \right)}}.
\end{equation}
However, due to the non-negligible interaction between the system and the environment, the actual steady state corresponds to the MFG state \cite{10.1116/5.0073853, PhysRevLett.127.250601,10.1063/1.4718706,PhysRevE.106.054145}
\begin{gather}
  {\rho _S} = {\tau _S} - {\beta _1}\sum\limits_\mu  {{\tau _S}} S\left( {{\omega _\mu }} \right) \hfill \notag \\
   \times \left( {V_1^\dag \left( {{\omega _\mu }} \right){V_1}\left( {{\omega _\mu }} \right) - Tr\left( {{\tau _S}V_1^\dag \left( {{\omega _\mu }} \right){V_1}\left( {{\omega _\mu }} \right)} \right)} \right) \hfill \notag \\
   - \sum\limits_{\mu  \ne \nu } {\frac{1}{{{\omega _\mu } - {\omega _\nu }}}S\left( {{\omega _\nu }} \right)}  \hfill \notag \\
   \times \left( {\left[ {V_1^\dag \left( {{\omega _\mu }} \right),{V_1}\left( {{\omega _\nu }} \right){\tau _S}} \right] + \left[ {{\tau _S}V_1^\dag \left( {{\omega _\nu }} \right),{V_1}\left( {{\omega _\mu }} \right)} \right]} \right) \hfill \notag \\
   - \sum\limits_\mu  {\left[ {{V_1}\left( {{\omega _\mu }} \right),{\tau _S}V_1^\dag \left( {{\omega _\mu }} \right)} \right]{{\left. {\frac{{dS\left( \omega  \right)}}{{d\omega }}} \right|}_{\omega  = {\omega _\mu }}}} , \hfill
\end{gather}
where the correction, in this case, precisely matches the second-order correction Eqs. (\ref{B4},\ref{B5}) obtained from the BR master equation. When $\gamma_1\rightarrow 0$, the correction can be negligible, and the steady state recovers the Gibbs state $\tau_S$. In the second case where $T_1=T_2=T$, the system also reaches an equilibrium state corresponding to the other MFG steady state. For both cases, the second-order correction exhibit excellent agreement, as illustrated in Fig. \ref{fig3}.


\section{Heat currents}
\label{4} 
We now focus on the non-equilibrium steady state where ${T_2} \ne {T_1}$ and ${\gamma _1} \ne 0, {\gamma _2} \ne 0.$ In this scenario, the heat spontaneously flows from the high-temperature reservoir to the low-temperature reservoir through the TLAs system. Our primary interest lies in the steady-state heat current $\mathcal{J}_j=Tr((H_S+H_{LS})\mathcal{L}_j)$.



\begin{figure}[h]
\centering\includegraphics[width=9cm]{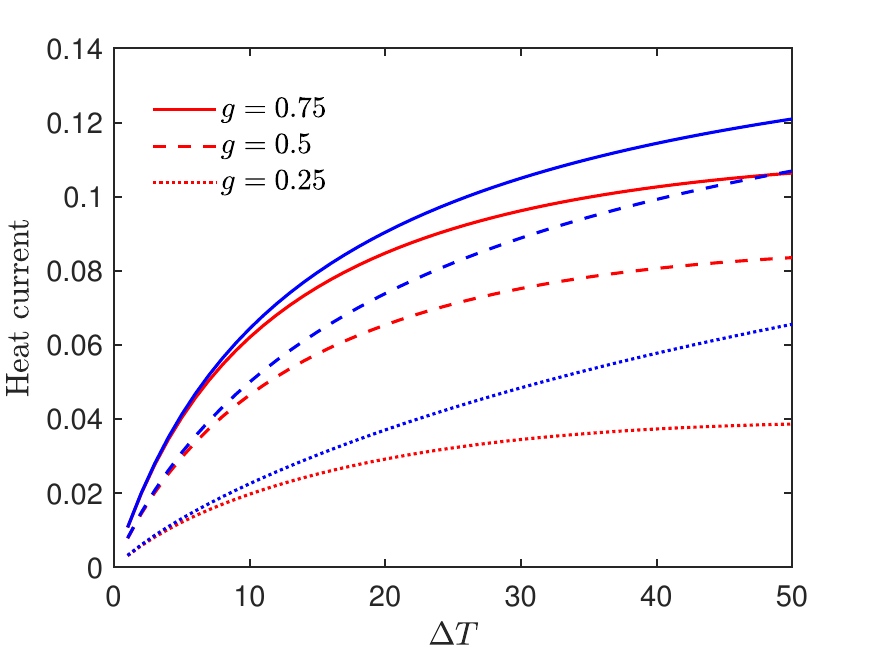}
\caption{The heat currents ${\mathcal{J}_{1}}$ vs the temperature difference $\Delta T$. Here we set $T_{2}=5,T_{1}=5+\Delta T,\varepsilon_1=5,\varepsilon_2=4,\gamma _{1}=\gamma _{2}=0.01,\omega _{D}=50.$ The blue lines represent the heat current obtained by the Lindblad master equation, and the red lines represent the heat current obtained by the BR master equation. For the solid, dashed, dotted lines, $g=0.75$, $0.5$, $0.25$, respectively. }
\label{fig1}
\end{figure}

\begin{figure}[h]
\centering\includegraphics[width=9cm]{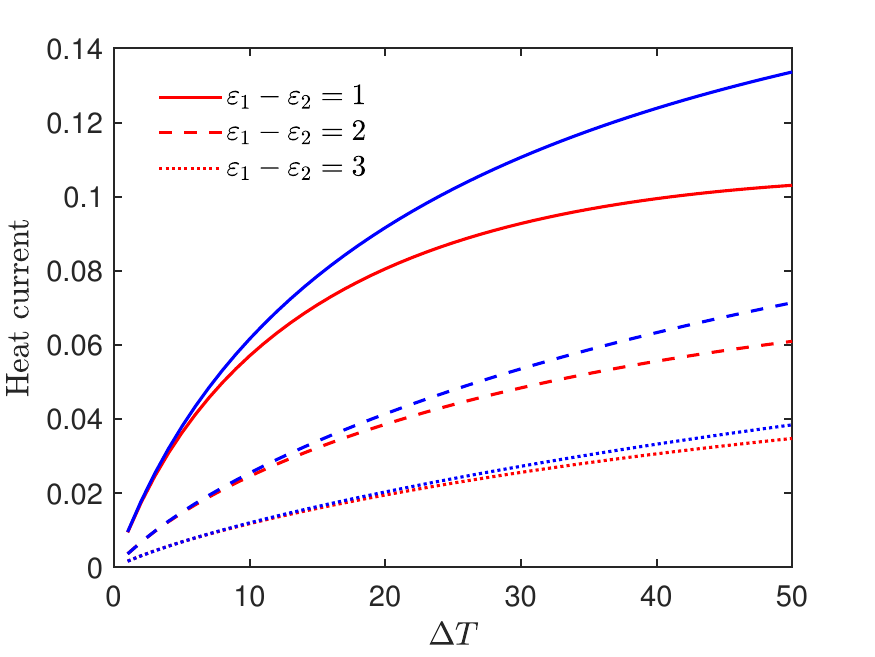}
\caption{The heat currents ${\mathcal{J}_{1}}$ vs the temperature difference $\Delta T$. Here we set $T_{2}=5,T_{1}=5+\Delta T,\varepsilon_1+\varepsilon_2=10,g=0.5,\gamma _{1}=\gamma _{2}=0.01,\omega _{D}=50.$ The blue lines represent the heat current obtained by the Lindblad master equation, and the red lines represent the heat current obtained by the BR master equation. For the solid, dashed, dotted lines, $(\varepsilon_1,\varepsilon_2)=(5.5,4.5)$, $(6,4)$, $(6.5,3.5)$, respectively. }
\label{fig2}
\end{figure}

Based on the BR master equation, one can find that $[H_{S}, H_{LS}] \ne 0$, but the conservation law remains valid i.e.,
\begin{eqnarray}{\mathcal{J}_1} + {\mathcal{J}_2} = Tr\left( {\left( {{H_S} + {H_{LS}}} \right)\left( {{\mathcal{L}_1}\left( {{\rho _S}} \right) + {\mathcal{L}_2}\left( {{\rho _S}} \right)} \right)} \right) \notag\\
= iTr\left( {\left( {{H_S} + {H_{LS}}} \right)\left[ {{H_S} + {H_{LS}},{\rho _S}} \right]} \right) = 0.
\end{eqnarray}
In Fig. \ref{fig1} and Fig. \ref{fig2}, we plot the heat currents obtained from the BR master equation and the Lindblad master equation, respectively. It is observed that the heat current for the BR master equation is always smaller than that derived from the Lindblad master equation. This reflects the suppressive effect of the non-secular terms on the heat current. In particular, the suppression magnitude increases with the temperature difference when all other parameters remain constant.

In addition, Fig. \ref{fig1} illustrates that the magnitude of suppression in the heat current is negatively correlated with $g$ under fixed parameters. Similarly, Fig. \ref{fig2} shows that while keeping $\varepsilon_1+\varepsilon_2$ constant, the magnitude of suppression in the heat current is negatively correlated with $\varepsilon_1-\varepsilon_2$. Namely, with the eigenenergy $\beta$ corresponding to the states $\left| {{s_1}} \right\rangle $ and $\left| {{s_4}} \right\rangle $ unchanged, the suppression is negatively correlated with the eigenenergy $\alpha$ corresponding to $\left| {{s_2}} \right\rangle $ and $\left| {{s_3}} \right\rangle $. This also implies that when $\varepsilon_1=\varepsilon_2=\varepsilon$ (i.e., in the symmetric case), the difference between the heat currents predicted by the Lindblad master equation and the BR master equation should be the largest.
However, we find that in the symmetric case, the Lindblad master equation loses its validity with a small $g\ll1$. This is because, under these conditions, the secular approximation---which is a fundamental assumption in deriving the Lindblad master equation---fails to hold: ${\omega _1} - {\omega _2} = 2\alpha  = 2g \sim \tau _R^{ - 1}$. Although we can still compute the steady-state solution and steady-state heat current for the Lindblad master equation in this scenario, these results lack physical meaning and exhibit unphysical behavior, such as the steady-state heat current not approaching zero as $g$ tends to zero, a more detailed discussion can be found in Appendix \ref{appendixC}.
\begin{figure}[h]
\centering\includegraphics[width=9cm]{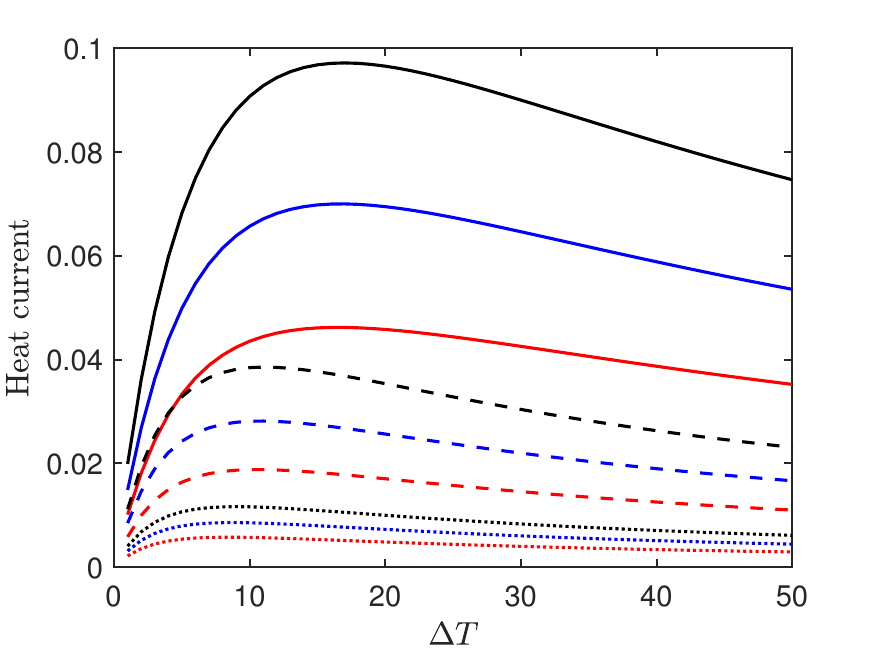}
\caption{The heat currents ${\mathcal{J}_{1}}$ versus the temperature difference $\Delta T$. Here we set $T_{2}=5,T_{1}=5+\Delta T,\gamma _{1}=\gamma _{2}=0.01,\omega _{D}=50.$ For the red, blue, and black lines, the parameters are $(\varepsilon_1,\varepsilon_2)=(4,4)$, $(5,5)$, and $(6,6)$ respectively, while for the solid, dashed, and dotted lines, $g=0.2$, $0.1$, and $0.05$ respectively.}
\label{fig4}
\end{figure}

\begin{figure}[h]
\centering\includegraphics[width=9cm]{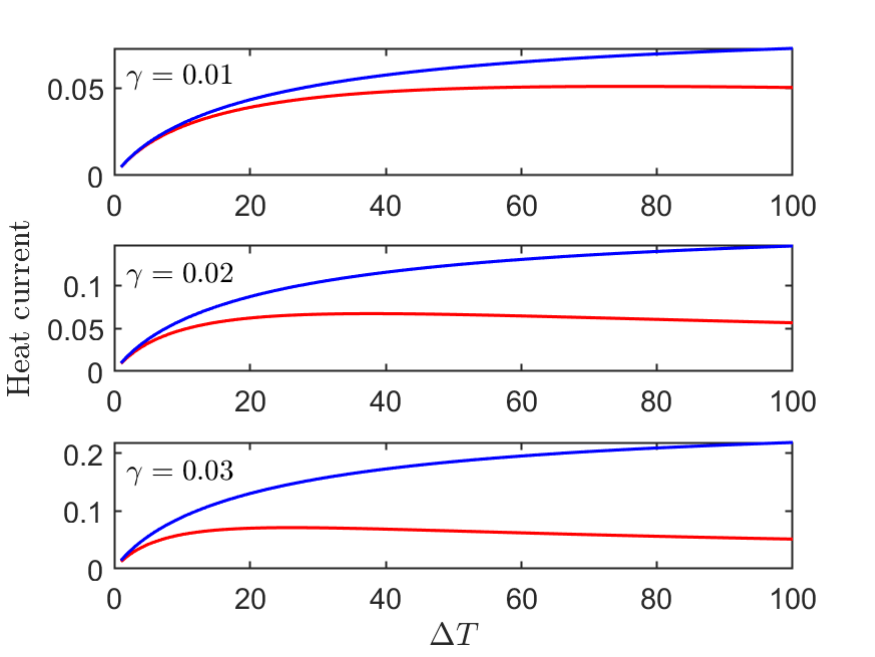}
\caption{The heat currents ${\mathcal{J}_{1}}$ vs the temperature difference $\Delta T$. Here we set $T_{2}=5,T_{1}=5+\Delta T,\varepsilon_1=4,\varepsilon_2=3,g=0.5,\omega _{D}=50.$ For the first inset, $\gamma _{1}=\gamma _{2}=0.01$; for the second inset, $\gamma _{1}=\gamma _{2}=0.02$; for the third inset, $\gamma _{1}=\gamma _{2}=0.03$. The blue and red lines represent the heat currents obtained using the Lindblad and BR master equations, respectively. }
\label{fig5}
\end{figure}

Under the symmetric conditions, the BR master equation remains valid and physically meaningful. We plot the heat currents in Fig. \ref{fig4}, which shows that the steady-state heat current significantly decreases as $g$ decreases. It indicates that in the symmetric case, the heat current obtained from the BR master equation vanishes as $g$ tends to zero (see Appendix \ref{appendixC} for a detailed derivation). Furthermore, one can find that the increase in $\varepsilon$ amplifies the heat current. As the temperature difference increases, the heat current initially rises and then decreases, exhibiting the NDTC effect. In fact, the NDTC exists in the asymmetric case as well. It only requires a larger temperature difference or a larger $\gamma$ compared to the symmetric case as shown in Fig. \ref{fig5}. As $\gamma$ increases, the NDTC effect becomes more pronounced, which is caused by BR's correction to the Lindblad master equation.

Considering the heat current based on the BR master equation, we can expand the density matrix to the second order as ${\rho ^{BR}} = {\rho ^{L}} + {\lambda ^2}{\rho ^{\left( 2 \right)}} + ...$. Thus the heat current can be given as
\begin{eqnarray}
&&{\mathcal{J}^{BR}_j} = Tr {(\left( {{H_S} + H_{LS}^{BR}} \right){\mathcal{L}^{BR}_j}\left( {{\rho ^{BR}}} \right))}\notag\\
&&={\mathcal{J}^{L}_j} + {\lambda ^2}Tr\left( {\left( {{H_S} + H_{LS}^{BR}} \right){\mathcal{L}^{BR}_j}\left( {{\rho ^{\left( 2 \right)}} + ...} \right)} \right),
\label{BRcurrent}
\end{eqnarray}
where 
\begin{eqnarray}{\mathcal{J}^{L}_j}&& =
Tr\left( {\left( {{H_S} + H_{LS}^{BR}} \right){\mathcal{L}^{BR}_j}\left( {{\rho ^{L}}} \right)} \right) \notag\\
&&= Tr\left( {\left( {{H_S} + H_{LS}^{L}} \right){\mathcal{L}^{L}_j}\left( {{\rho ^{L}}} \right)} \right).
\end{eqnarray}
Eq. (\ref{BRcurrent}) shows that the corrections to the heat current arise entirely from the second and higher-order contributions of the steady-state solution. Consequently, if only the zeroth-order approximation is considered, the magnitude of the heat current predicted by the BR master equation does not differ from that given by the Lindblad master equation despite the explicit inclusion of non-secular terms in the former.

In our model, it can be shown that the NDTC emerges exclusively in the BR master equation, which can be verified as follows. The heat current derived from the Lindblad master equation is given by
\begin{equation}
{\mathcal{J}_{1}^{L}}=\sum_{i=1}^{2}A_{i}\left( {{\bar{n}_{1}}\left( {
\omega _{i}}\right) -{\bar{n}_{2}}\left( {\omega _{i}}\right) }\right)
\left( {{\omega _{i}}+{\delta _{i}}}\right),
\label{lindcurrent}
\end{equation}
with 
\begin{gather}
A_{1}=2{\sin ^{2}}{\phi _{+}}{\cos ^{2}}{\phi _{-}}\frac{{{J_{1}}\left( {{%
\omega _{1}}}\right) {J_{2}}\left( {{\omega _{1}}}\right) }}{{{\widetilde G}_ + }\left( {{\omega _1}} \right) + {{\widetilde G}_ + }\left( { - {\omega _1}} \right)},\hfill \\
A_{2}=2{\sin ^{2}}{\phi _{-}}{\cos ^{2}}{\phi _{+}}\frac{{{J_{1}}\left( {{%
\omega _{2}}}\right) {J_{2}}\left( {{\omega _{2}}}\right) }}{{\widetilde G_ + }\left( {{\omega _2}} \right) + {\widetilde G_ + }\left( { - {\omega _2}} \right)},\hfill
\end{gather}
and $\delta_i$ are second-order corrections to the eigenfrequency $\omega_i$ introduced by the lamb shift. We can prove that when $T_2$ is fixed and $T_1>T_2$,
$\partial \mathcal{J}_1^{L}/\partial {T_1} > 0$, which shows that the heat current given by Lindblad's master equation does not exhibit the NDTC effect. 

\section{Conclusions}
\label{5}
We have studied the heat currents through a coupled TLA system based on the BR master equation under the Born-Markov approximation.  Using a perturbative approach, we yield analytical expressions of the steady-state density matrix at zeroth and second order. The zeroth-order solution coincides precisely with the steady-state density matrix of the Lindblad master equation, and the second-order solution aligns with the mean force Gibbs state when an equilibrium state exists in the regime.
 By studying the heat current in our regime, we discovered that the non-secular terms exert a suppressive effect on the heat current, and the magnitude of this suppression is negatively correlated with $g$ and $\varepsilon_1-\varepsilon_2$.
In particular, we identified the emergence of NDTC, which is a key nonlinear transport feature, in specific parameter regimes within the BR formalism. This phenomenon, absent in the Lindblad description, manifests as a decrease in heat current with increasing temperature bias, highlighting the essential role of coherence and quantum correlations in shaping thermal response. Furthermore, it shows that in the symmetric case, the Lindblad master equation yields unphysical results as $g$ approaches zero, whereas the BR master equation remains valid regardless of the coupling strength. 

In summary, this work demonstrates the critical importance of non-secular contributions in capturing realistic quantum thermal transport phenomena, particularly the emergence of NDTC. Additionally, we also find that the concurrence of assistance and the linear entropy have behaviors similar to those of the NDTC, which is not provided here. Our findings highlight the benefits of the Bloch-Redfield formalism in describing the dynamics of open quantum systems beyond the secular approximation, thereby paving the way for more accurate modeling and control of heat flow in quantum thermal devices.

\section*{Acknowledgments}
 This work was supported by the National Natural Science Foundation of China under Grants Nos. 12175029.
\section*{References}
\bibliography{book}

\begin{thebibliography}{74}%
\makeatletter
\providecommand \@ifxundefined [1]{%
 \@ifx{#1\undefined}
}%
\providecommand \@ifnum [1]{%
 \ifnum #1\expandafter \@firstoftwo
 \else \expandafter \@secondoftwo
 \fi
}%
\providecommand \@ifx [1]{%
 \ifx #1\expandafter \@firstoftwo
 \else \expandafter \@secondoftwo
 \fi
}%
\providecommand \natexlab [1]{#1}%
\providecommand \enquote  [1]{``#1''}%
\providecommand \bibnamefont  [1]{#1}%
\providecommand \bibfnamefont [1]{#1}%
\providecommand \citenamefont [1]{#1}%
\providecommand \href@noop [0]{\@secondoftwo}%
\providecommand \href [0]{\begingroup \@sanitize@url \@href}%
\providecommand \@href[1]{\@@startlink{#1}\@@href}%
\providecommand \@@href[1]{\endgroup#1\@@endlink}%
\providecommand \@sanitize@url [0]{\catcode `\\12\catcode `\$12\catcode
  `\&12\catcode `\#12\catcode `\^12\catcode `\_12\catcode `\%12\relax}%
\providecommand \@@startlink[1]{}%
\providecommand \@@endlink[0]{}%
\providecommand \url  [0]{\begingroup\@sanitize@url \@url }%
\providecommand \@url [1]{\endgroup\@href {#1}{\urlprefix }}%
\providecommand \urlprefix  [0]{URL }%
\providecommand \Eprint [0]{\href }%
\providecommand \doibase [0]{https://doi.org/}%
\providecommand \selectlanguage [0]{\@gobble}%
\providecommand \bibinfo  [0]{\@secondoftwo}%
\providecommand \bibfield  [0]{\@secondoftwo}%
\providecommand \translation [1]{[#1]}%
\providecommand \BibitemOpen [0]{}%
\providecommand \bibitemStop [0]{}%
\providecommand \bibitemNoStop [0]{.\EOS\space}%
\providecommand \EOS [0]{\spacefactor3000\relax}%
\providecommand \BibitemShut  [1]{\csname bibitem#1\endcsname}%
\let\auto@bib@innerbib\@empty
\bibitem [{\citenamefont {Feldmann}\ and\ \citenamefont
  {Kosloff}(2000)}]{PhysRevE.61.4774}%
  \BibitemOpen
  \bibfield  {author} {\bibinfo {author} {\bibfnamefont {T.}~\bibnamefont
  {Feldmann}}\ and\ \bibinfo {author} {\bibfnamefont {R.}~\bibnamefont
  {Kosloff}},\ }\bibfield  {title} {\bibinfo {title} {Performance of discrete
  heat engines and heat pumps in finite time},\ }\href
  {https://doi.org/10.1103/PhysRevE.61.4774} {\bibfield  {journal} {\bibinfo
  {journal} {Phys. Rev. E}\ }\textbf {\bibinfo {volume} {61}},\ \bibinfo
  {pages} {4774} (\bibinfo {year} {2000})}\BibitemShut {NoStop}%
\bibitem [{\citenamefont {Segal}\ and\ \citenamefont
  {Nitzan}(2006)}]{PhysRevE.73.026109}%
  \BibitemOpen
  \bibfield  {author} {\bibinfo {author} {\bibfnamefont {D.}~\bibnamefont
  {Segal}}\ and\ \bibinfo {author} {\bibfnamefont {A.}~\bibnamefont {Nitzan}},\
  }\bibfield  {title} {\bibinfo {title} {Molecular heat pump},\ }\href
  {https://doi.org/10.1103/PhysRevE.73.026109} {\bibfield  {journal} {\bibinfo
  {journal} {Phys. Rev. E}\ }\textbf {\bibinfo {volume} {73}},\ \bibinfo
  {pages} {026109} (\bibinfo {year} {2006})}\BibitemShut {NoStop}%
\bibitem [{\citenamefont {Geva}\ and\ \citenamefont
  {Kosloff}(1992)}]{doi:10.1063/1.461951}%
  \BibitemOpen
  \bibfield  {author} {\bibinfo {author} {\bibfnamefont {E.}~\bibnamefont
  {Geva}}\ and\ \bibinfo {author} {\bibfnamefont {R.}~\bibnamefont {Kosloff}},\
  }\bibfield  {title} {\bibinfo {title} {A quantum‐mechanical heat engine
  operating in finite time. a model consisting of spin‐1/2 systems as the
  working fluid},\ }\href {https://doi.org/10.1063/1.461951} {\bibfield
  {journal} {\bibinfo  {journal} {The Journal of Chemical Physics}\ }\textbf
  {\bibinfo {volume} {96}},\ \bibinfo {pages} {3054} (\bibinfo {year}
  {1992})},\ \Eprint {https://arxiv.org/abs/https://doi.org/10.1063/1.461951}
  {https://doi.org/10.1063/1.461951} \BibitemShut {NoStop}%
\bibitem [{\citenamefont {Ono}\ \emph {et~al.}(2020)\citenamefont {Ono},
  \citenamefont {Shevchenko}, \citenamefont {Mori}, \citenamefont {Moriyama},\
  and\ \citenamefont {Nori}}]{PhysRevLett.125.166802}%
  \BibitemOpen
  \bibfield  {author} {\bibinfo {author} {\bibfnamefont {K.}~\bibnamefont
  {Ono}}, \bibinfo {author} {\bibfnamefont {S.~N.}\ \bibnamefont {Shevchenko}},
  \bibinfo {author} {\bibfnamefont {T.}~\bibnamefont {Mori}}, \bibinfo {author}
  {\bibfnamefont {S.}~\bibnamefont {Moriyama}},\ and\ \bibinfo {author}
  {\bibfnamefont {F.}~\bibnamefont {Nori}},\ }\bibfield  {title} {\bibinfo
  {title} {Analog of a quantum heat engine using a single-spin qubit},\ }\href
  {https://doi.org/10.1103/PhysRevLett.125.166802} {\bibfield  {journal}
  {\bibinfo  {journal} {Phys. Rev. Lett.}\ }\textbf {\bibinfo {volume} {125}},\
  \bibinfo {pages} {166802} (\bibinfo {year} {2020})}\BibitemShut {NoStop}%
\bibitem [{\citenamefont {Naghiloo}\ \emph {et~al.}(2020)\citenamefont
  {Naghiloo}, \citenamefont {Tan}, \citenamefont {Harrington}, \citenamefont
  {Alonso}, \citenamefont {Lutz}, \citenamefont {Romito},\ and\ \citenamefont
  {Murch}}]{PhysRevLett.124.110604}%
  \BibitemOpen
  \bibfield  {author} {\bibinfo {author} {\bibfnamefont {M.}~\bibnamefont
  {Naghiloo}}, \bibinfo {author} {\bibfnamefont {D.}~\bibnamefont {Tan}},
  \bibinfo {author} {\bibfnamefont {P.~M.}\ \bibnamefont {Harrington}},
  \bibinfo {author} {\bibfnamefont {J.~J.}\ \bibnamefont {Alonso}}, \bibinfo
  {author} {\bibfnamefont {E.}~\bibnamefont {Lutz}}, \bibinfo {author}
  {\bibfnamefont {A.}~\bibnamefont {Romito}},\ and\ \bibinfo {author}
  {\bibfnamefont {K.~W.}\ \bibnamefont {Murch}},\ }\bibfield  {title} {\bibinfo
  {title} {Heat and work along individual trajectories of a quantum bit},\
  }\href {https://doi.org/10.1103/PhysRevLett.124.110604} {\bibfield  {journal}
  {\bibinfo  {journal} {Phys. Rev. Lett.}\ }\textbf {\bibinfo {volume} {124}},\
  \bibinfo {pages} {110604} (\bibinfo {year} {2020})}\BibitemShut {NoStop}%
\bibitem [{\citenamefont {Josefsson}\ and\ \citenamefont
  {Leijnse}(2020)}]{PhysRevB.101.081408}%
  \BibitemOpen
  \bibfield  {author} {\bibinfo {author} {\bibfnamefont {M.}~\bibnamefont
  {Josefsson}}\ and\ \bibinfo {author} {\bibfnamefont {M.}~\bibnamefont
  {Leijnse}},\ }\bibfield  {title} {\bibinfo {title} {Double quantum-dot engine
  fueled by entanglement between electron spins},\ }\href
  {https://doi.org/10.1103/PhysRevB.101.081408} {\bibfield  {journal} {\bibinfo
   {journal} {Phys. Rev. B}\ }\textbf {\bibinfo {volume} {101}},\ \bibinfo
  {pages} {081408} (\bibinfo {year} {2020})}\BibitemShut {NoStop}%
\bibitem [{\citenamefont {Dann}\ \emph {et~al.}(2020)\citenamefont {Dann},
  \citenamefont {Kosloff},\ and\ \citenamefont {Salamon}}]{e22111255}%
  \BibitemOpen
  \bibfield  {author} {\bibinfo {author} {\bibfnamefont {R.}~\bibnamefont
  {Dann}}, \bibinfo {author} {\bibfnamefont {R.}~\bibnamefont {Kosloff}},\ and\
  \bibinfo {author} {\bibfnamefont {P.}~\bibnamefont {Salamon}},\ }\bibfield
  {title} {\bibinfo {title} {Quantum finite-time thermodynamics: Insight from a
  single qubit engine},\ }\bibfield  {journal} {\bibinfo  {journal} {Entropy}\
  }\textbf {\bibinfo {volume} {22}},\ \href {https://doi.org/10.3390/e22111255}
  {10.3390/e22111255} (\bibinfo {year} {2020})\BibitemShut {NoStop}%
\bibitem [{\citenamefont {Maillet}\ \emph {et~al.}(2019)\citenamefont
  {Maillet}, \citenamefont {Erdman}, \citenamefont {Cavina}, \citenamefont
  {Bhandari}, \citenamefont {Mannila}, \citenamefont {Peltonen}, \citenamefont
  {Mari}, \citenamefont {Taddei}, \citenamefont {Jarzynski}, \citenamefont
  {Giovannetti},\ and\ \citenamefont {Pekola}}]{PhysRevLett.122.150604}%
  \BibitemOpen
  \bibfield  {author} {\bibinfo {author} {\bibfnamefont {O.}~\bibnamefont
  {Maillet}}, \bibinfo {author} {\bibfnamefont {P.~A.}\ \bibnamefont {Erdman}},
  \bibinfo {author} {\bibfnamefont {V.}~\bibnamefont {Cavina}}, \bibinfo
  {author} {\bibfnamefont {B.}~\bibnamefont {Bhandari}}, \bibinfo {author}
  {\bibfnamefont {E.~T.}\ \bibnamefont {Mannila}}, \bibinfo {author}
  {\bibfnamefont {J.~T.}\ \bibnamefont {Peltonen}}, \bibinfo {author}
  {\bibfnamefont {A.}~\bibnamefont {Mari}}, \bibinfo {author} {\bibfnamefont
  {F.}~\bibnamefont {Taddei}}, \bibinfo {author} {\bibfnamefont
  {C.}~\bibnamefont {Jarzynski}}, \bibinfo {author} {\bibfnamefont
  {V.}~\bibnamefont {Giovannetti}},\ and\ \bibinfo {author} {\bibfnamefont
  {J.~P.}\ \bibnamefont {Pekola}},\ }\bibfield  {title} {\bibinfo {title}
  {Optimal probabilistic work extraction beyond the free energy difference with
  a single-electron device},\ }\href
  {https://doi.org/10.1103/PhysRevLett.122.150604} {\bibfield  {journal}
  {\bibinfo  {journal} {Phys. Rev. Lett.}\ }\textbf {\bibinfo {volume} {122}},\
  \bibinfo {pages} {150604} (\bibinfo {year} {2019})}\BibitemShut {NoStop}%
\bibitem [{\citenamefont {Souza}\ \emph {et~al.}(2022)\citenamefont {Souza},
  \citenamefont {Manzano}, \citenamefont {Fazio},\ and\ \citenamefont
  {Iemini}}]{PhysRevE.106.014143}%
  \BibitemOpen
  \bibfield  {author} {\bibinfo {author} {\bibfnamefont {L.~d.~S.}\
  \bibnamefont {Souza}}, \bibinfo {author} {\bibfnamefont {G.}~\bibnamefont
  {Manzano}}, \bibinfo {author} {\bibfnamefont {R.}~\bibnamefont {Fazio}},\
  and\ \bibinfo {author} {\bibfnamefont {F.}~\bibnamefont {Iemini}},\
  }\bibfield  {title} {\bibinfo {title} {Collective effects on the performance
  and stability of quantum heat engines},\ }\href
  {https://doi.org/10.1103/PhysRevE.106.014143} {\bibfield  {journal} {\bibinfo
   {journal} {Phys. Rev. E}\ }\textbf {\bibinfo {volume} {106}},\ \bibinfo
  {pages} {014143} (\bibinfo {year} {2022})}\BibitemShut {NoStop}%
\bibitem [{\citenamefont {Colla}\ and\ \citenamefont
  {Breuer}(2024)}]{Colla_2025}%
  \BibitemOpen
  \bibfield  {author} {\bibinfo {author} {\bibfnamefont {A.}~\bibnamefont
  {Colla}}\ and\ \bibinfo {author} {\bibfnamefont {H.-P.}\ \bibnamefont
  {Breuer}},\ }\bibfield  {title} {\bibinfo {title} {Thermodynamic roles of
  quantum environments: from heat baths to work reservoirs},\ }\href
  {https://doi.org/10.1088/2058-9565/ad98be} {\bibfield  {journal} {\bibinfo
  {journal} {Quantum Science and Technology}\ }\textbf {\bibinfo {volume}
  {10}},\ \bibinfo {pages} {015047} (\bibinfo {year} {2024})}\BibitemShut
  {NoStop}%
\bibitem [{\citenamefont {Palao}\ \emph {et~al.}(2001)\citenamefont {Palao},
  \citenamefont {Kosloff},\ and\ \citenamefont {Gordon}}]{PhysRevE.64.056130}%
  \BibitemOpen
  \bibfield  {author} {\bibinfo {author} {\bibfnamefont {J.~P.}\ \bibnamefont
  {Palao}}, \bibinfo {author} {\bibfnamefont {R.}~\bibnamefont {Kosloff}},\
  and\ \bibinfo {author} {\bibfnamefont {J.~M.}\ \bibnamefont {Gordon}},\
  }\bibfield  {title} {\bibinfo {title} {Quantum thermodynamic cooling cycle},\
  }\href {https://doi.org/10.1103/PhysRevE.64.056130} {\bibfield  {journal}
  {\bibinfo  {journal} {Phys. Rev. E}\ }\textbf {\bibinfo {volume} {64}},\
  \bibinfo {pages} {056130} (\bibinfo {year} {2001})}\BibitemShut {NoStop}%
\bibitem [{\citenamefont {Feldmann}\ \emph {et~al.}(1996)\citenamefont
  {Feldmann}, \citenamefont {Geva}, \citenamefont {Kosloff},\ and\
  \citenamefont {Salamon}}]{doi:10.1119/1.18197}%
  \BibitemOpen
  \bibfield  {author} {\bibinfo {author} {\bibfnamefont {T.}~\bibnamefont
  {Feldmann}}, \bibinfo {author} {\bibfnamefont {E.}~\bibnamefont {Geva}},
  \bibinfo {author} {\bibfnamefont {R.}~\bibnamefont {Kosloff}},\ and\ \bibinfo
  {author} {\bibfnamefont {P.}~\bibnamefont {Salamon}},\ }\bibfield  {title}
  {\bibinfo {title} {Heat engines in finite time governed by master
  equations},\ }\href {https://doi.org/10.1119/1.18197} {\bibfield  {journal}
  {\bibinfo  {journal} {American Journal of Physics}\ }\textbf {\bibinfo
  {volume} {64}},\ \bibinfo {pages} {485} (\bibinfo {year} {1996})},\ \Eprint
  {https://arxiv.org/abs/https://doi.org/10.1119/1.18197}
  {https://doi.org/10.1119/1.18197} \BibitemShut {NoStop}%
\bibitem [{\citenamefont {Skrzypczyk}\ \emph {et~al.}(2011)\citenamefont
  {Skrzypczyk}, \citenamefont {Brunner}, \citenamefont {Linden},\ and\
  \citenamefont {Popescu}}]{Skrzypczyk_2011}%
  \BibitemOpen
  \bibfield  {author} {\bibinfo {author} {\bibfnamefont {P.}~\bibnamefont
  {Skrzypczyk}}, \bibinfo {author} {\bibfnamefont {N.}~\bibnamefont {Brunner}},
  \bibinfo {author} {\bibfnamefont {N.}~\bibnamefont {Linden}},\ and\ \bibinfo
  {author} {\bibfnamefont {S.}~\bibnamefont {Popescu}},\ }\bibfield  {title}
  {\bibinfo {title} {The smallest refrigerators can reach maximal efficiency},\
  }\href {https://doi.org/10.1088/1751-8113/44/49/492002} {\bibfield  {journal}
  {\bibinfo  {journal} {Journal of Physics A: Mathematical and Theoretical}\
  }\textbf {\bibinfo {volume} {44}},\ \bibinfo {pages} {492002} (\bibinfo
  {year} {2011})}\BibitemShut {NoStop}%
\bibitem [{\citenamefont {Yu}\ and\ \citenamefont {Zhu}(2014)}]{yu2014re}%
  \BibitemOpen
  \bibfield  {author} {\bibinfo {author} {\bibfnamefont {C.-s.}\ \bibnamefont
  {Yu}}\ and\ \bibinfo {author} {\bibfnamefont {Q.-y.}\ \bibnamefont {Zhu}},\
  }\bibfield  {title} {\bibinfo {title} {Re-examining the self-contained
  quantum refrigerator in the strong-coupling regime},\ }\href@noop {}
  {\bibfield  {journal} {\bibinfo  {journal} {Physical Review E}\ }\textbf
  {\bibinfo {volume} {90}},\ \bibinfo {pages} {052142} (\bibinfo {year}
  {2014})}\BibitemShut {NoStop}%
\bibitem [{\citenamefont {Yu}\ \emph {et~al.}(2019)\citenamefont {Yu},
  \citenamefont {Guo},\ and\ \citenamefont {Liu}}]{yu2019quantum}%
  \BibitemOpen
  \bibfield  {author} {\bibinfo {author} {\bibfnamefont {C.}~\bibnamefont
  {Yu}}, \bibinfo {author} {\bibfnamefont {B.}~\bibnamefont {Guo}},\ and\
  \bibinfo {author} {\bibfnamefont {T.}~\bibnamefont {Liu}},\ }\bibfield
  {title} {\bibinfo {title} {Quantum self-contained refrigerator in terms of
  the cavity quantum electrodynamics in the weak internal-coupling regime.},\
  }\href@noop {} {\bibfield  {journal} {\bibinfo  {journal} {Optics Express}\
  }\textbf {\bibinfo {volume} {27}},\ \bibinfo {pages} {6863} (\bibinfo {year}
  {2019})}\BibitemShut {NoStop}%
\bibitem [{\citenamefont {Buffoni}\ \emph {et~al.}(2019)\citenamefont
  {Buffoni}, \citenamefont {Solfanelli}, \citenamefont {Verrucchi},
  \citenamefont {Cuccoli},\ and\ \citenamefont
  {Campisi}}]{PhysRevLett.122.070603}%
  \BibitemOpen
  \bibfield  {author} {\bibinfo {author} {\bibfnamefont {L.}~\bibnamefont
  {Buffoni}}, \bibinfo {author} {\bibfnamefont {A.}~\bibnamefont {Solfanelli}},
  \bibinfo {author} {\bibfnamefont {P.}~\bibnamefont {Verrucchi}}, \bibinfo
  {author} {\bibfnamefont {A.}~\bibnamefont {Cuccoli}},\ and\ \bibinfo {author}
  {\bibfnamefont {M.}~\bibnamefont {Campisi}},\ }\bibfield  {title} {\bibinfo
  {title} {Quantum measurement cooling},\ }\href
  {https://doi.org/10.1103/PhysRevLett.122.070603} {\bibfield  {journal}
  {\bibinfo  {journal} {Phys. Rev. Lett.}\ }\textbf {\bibinfo {volume} {122}},\
  \bibinfo {pages} {070603} (\bibinfo {year} {2019})}\BibitemShut {NoStop}%
\bibitem [{\citenamefont {Soldati}\ \emph {et~al.}(2022)\citenamefont
  {Soldati}, \citenamefont {Dasari}, \citenamefont {Wrachtrup},\ and\
  \citenamefont {Lutz}}]{PhysRevLett.129.030601}%
  \BibitemOpen
  \bibfield  {author} {\bibinfo {author} {\bibfnamefont {R.~R.}\ \bibnamefont
  {Soldati}}, \bibinfo {author} {\bibfnamefont {D.~B.~R.}\ \bibnamefont
  {Dasari}}, \bibinfo {author} {\bibfnamefont {J.}~\bibnamefont {Wrachtrup}},\
  and\ \bibinfo {author} {\bibfnamefont {E.}~\bibnamefont {Lutz}},\ }\bibfield
  {title} {\bibinfo {title} {Thermodynamics of a minimal algorithmic cooling
  refrigerator},\ }\href {https://doi.org/10.1103/PhysRevLett.129.030601}
  {\bibfield  {journal} {\bibinfo  {journal} {Phys. Rev. Lett.}\ }\textbf
  {\bibinfo {volume} {129}},\ \bibinfo {pages} {030601} (\bibinfo {year}
  {2022})}\BibitemShut {NoStop}%
\bibitem [{\citenamefont {Bhandari}\ and\ \citenamefont
  {Jordan}(2022)}]{PhysRevResearch.4.033103}%
  \BibitemOpen
  \bibfield  {author} {\bibinfo {author} {\bibfnamefont {B.}~\bibnamefont
  {Bhandari}}\ and\ \bibinfo {author} {\bibfnamefont {A.~N.}\ \bibnamefont
  {Jordan}},\ }\bibfield  {title} {\bibinfo {title} {Continuous measurement
  boosted adiabatic quantum thermal machines},\ }\href
  {https://doi.org/10.1103/PhysRevResearch.4.033103} {\bibfield  {journal}
  {\bibinfo  {journal} {Phys. Rev. Res.}\ }\textbf {\bibinfo {volume} {4}},\
  \bibinfo {pages} {033103} (\bibinfo {year} {2022})}\BibitemShut {NoStop}%
\bibitem [{\citenamefont {Elouard}\ \emph {et~al.}(2017)\citenamefont
  {Elouard}, \citenamefont {Herrera-Mart\'{\i}}, \citenamefont {Huard},\ and\
  \citenamefont {Auff\`eves}}]{PhysRevLett.118.260603}%
  \BibitemOpen
  \bibfield  {author} {\bibinfo {author} {\bibfnamefont {C.}~\bibnamefont
  {Elouard}}, \bibinfo {author} {\bibfnamefont {D.}~\bibnamefont
  {Herrera-Mart\'{\i}}}, \bibinfo {author} {\bibfnamefont {B.}~\bibnamefont
  {Huard}},\ and\ \bibinfo {author} {\bibfnamefont {A.}~\bibnamefont
  {Auff\`eves}},\ }\bibfield  {title} {\bibinfo {title} {Extracting work from
  quantum measurement in maxwell's demon engines},\ }\href
  {https://doi.org/10.1103/PhysRevLett.118.260603} {\bibfield  {journal}
  {\bibinfo  {journal} {Phys. Rev. Lett.}\ }\textbf {\bibinfo {volume} {118}},\
  \bibinfo {pages} {260603} (\bibinfo {year} {2017})}\BibitemShut {NoStop}%
\bibitem [{\citenamefont {Yanik}\ \emph {et~al.}(2022)\citenamefont {Yanik},
  \citenamefont {Bhandari}, \citenamefont {Manikandan},\ and\ \citenamefont
  {Jordan}}]{PhysRevA.106.042221}%
  \BibitemOpen
  \bibfield  {author} {\bibinfo {author} {\bibfnamefont {K.}~\bibnamefont
  {Yanik}}, \bibinfo {author} {\bibfnamefont {B.}~\bibnamefont {Bhandari}},
  \bibinfo {author} {\bibfnamefont {S.~K.}\ \bibnamefont {Manikandan}},\ and\
  \bibinfo {author} {\bibfnamefont {A.~N.}\ \bibnamefont {Jordan}},\ }\bibfield
   {title} {\bibinfo {title} {Thermodynamics of quantum measurement and
  maxwell's demon's arrow of time},\ }\href
  {https://doi.org/10.1103/PhysRevA.106.042221} {\bibfield  {journal} {\bibinfo
   {journal} {Phys. Rev. A}\ }\textbf {\bibinfo {volume} {106}},\ \bibinfo
  {pages} {042221} (\bibinfo {year} {2022})}\BibitemShut {NoStop}%
\bibitem [{\citenamefont {Jacobs}(2009)}]{PhysRevA.80.012322}%
  \BibitemOpen
  \bibfield  {author} {\bibinfo {author} {\bibfnamefont {K.}~\bibnamefont
  {Jacobs}},\ }\bibfield  {title} {\bibinfo {title} {Second law of
  thermodynamics and quantum feedback control: Maxwell's demon with weak
  measurements},\ }\href {https://doi.org/10.1103/PhysRevA.80.012322}
  {\bibfield  {journal} {\bibinfo  {journal} {Phys. Rev. A}\ }\textbf {\bibinfo
  {volume} {80}},\ \bibinfo {pages} {012322} (\bibinfo {year}
  {2009})}\BibitemShut {NoStop}%
\bibitem [{\citenamefont {Chan}\ \emph {et~al.}(2022)\citenamefont {Chan},
  \citenamefont {Huang}, \citenamefont {Lin}, \citenamefont {Ku}, \citenamefont
  {Chen}, \citenamefont {Chen},\ and\ \citenamefont
  {Chen}}]{PhysRevA.106.052201}%
  \BibitemOpen
  \bibfield  {author} {\bibinfo {author} {\bibfnamefont {F.-J.}\ \bibnamefont
  {Chan}}, \bibinfo {author} {\bibfnamefont {Y.-T.}\ \bibnamefont {Huang}},
  \bibinfo {author} {\bibfnamefont {J.-D.}\ \bibnamefont {Lin}}, \bibinfo
  {author} {\bibfnamefont {H.-Y.}\ \bibnamefont {Ku}}, \bibinfo {author}
  {\bibfnamefont {J.-S.}\ \bibnamefont {Chen}}, \bibinfo {author}
  {\bibfnamefont {H.-B.}\ \bibnamefont {Chen}},\ and\ \bibinfo {author}
  {\bibfnamefont {Y.-N.}\ \bibnamefont {Chen}},\ }\bibfield  {title} {\bibinfo
  {title} {Maxwell's two-demon engine under pure dephasing noise},\ }\href
  {https://doi.org/10.1103/PhysRevA.106.052201} {\bibfield  {journal} {\bibinfo
   {journal} {Phys. Rev. A}\ }\textbf {\bibinfo {volume} {106}},\ \bibinfo
  {pages} {052201} (\bibinfo {year} {2022})}\BibitemShut {NoStop}%
\bibitem [{\citenamefont {Breuer}\ \emph {et~al.}(2002)\citenamefont {Breuer},
  \citenamefont {Petruccione} \emph {et~al.}}]{breuer2002theory}%
  \BibitemOpen
  \bibfield  {author} {\bibinfo {author} {\bibfnamefont {H.-P.}\ \bibnamefont
  {Breuer}}, \bibinfo {author} {\bibfnamefont {F.}~\bibnamefont {Petruccione}},
  \emph {et~al.},\ }\href@noop {} {\emph {\bibinfo {title} {The theory of open
  quantum systems}}}\ (\bibinfo  {publisher} {Oxford University Press on
  Demand},\ \bibinfo {year} {2002})\BibitemShut {NoStop}%
\bibitem [{\citenamefont {Weiss}(2012)}]{weiss2012quantum}%
  \BibitemOpen
  \bibfield  {author} {\bibinfo {author} {\bibfnamefont {U.}~\bibnamefont
  {Weiss}},\ }\href@noop {} {\emph {\bibinfo {title} {Quantum dissipative
  systems}}}\ (\bibinfo  {publisher} {World Scientific},\ \bibinfo {year}
  {2012})\BibitemShut {NoStop}%
\bibitem [{\citenamefont {Carmichael}(2013)}]{carmichael2013statistical}%
  \BibitemOpen
  \bibfield  {author} {\bibinfo {author} {\bibfnamefont {H.~J.}\ \bibnamefont
  {Carmichael}},\ }\href@noop {} {\emph {\bibinfo {title} {Statistical methods
  in quantum optics 1: master equations and Fokker-Planck equations}}}\
  (\bibinfo  {publisher} {Springer Science \& Business Media},\ \bibinfo {year}
  {2013})\BibitemShut {NoStop}%
\bibitem [{\citenamefont {Schaller}(2014)}]{schaller2014non}%
  \BibitemOpen
  \bibfield  {author} {\bibinfo {author} {\bibfnamefont {G.}~\bibnamefont
  {Schaller}},\ }\bibfield  {title} {\bibinfo {title} {Non-equilibrium master
  equations},\ }\href@noop {} {\bibfield  {journal} {\bibinfo  {journal}
  {Lecture Notes TU Berlin, Berlin}\ } (\bibinfo {year} {2014})}\BibitemShut
  {NoStop}%
\bibitem [{\citenamefont {Lidar}(2019)}]{lidar2019lecture}%
  \BibitemOpen
  \bibfield  {author} {\bibinfo {author} {\bibfnamefont {D.~A.}\ \bibnamefont
  {Lidar}},\ }\bibfield  {title} {\bibinfo {title} {Lecture notes on the theory
  of open quantum systems},\ }\href@noop {} {\bibfield  {journal} {\bibinfo
  {journal} {arXiv preprint arXiv:1902.00967}\ } (\bibinfo {year}
  {2019})}\BibitemShut {NoStop}%
\bibitem [{\citenamefont {Oppenheim}\ and\ \citenamefont
  {Shuler}(1965)}]{PhysRev.138.B1007}%
  \BibitemOpen
  \bibfield  {author} {\bibinfo {author} {\bibfnamefont {I.}~\bibnamefont
  {Oppenheim}}\ and\ \bibinfo {author} {\bibfnamefont {K.~E.}\ \bibnamefont
  {Shuler}},\ }\bibfield  {title} {\bibinfo {title} {Master equations and
  markov processes},\ }\href {https://doi.org/10.1103/PhysRev.138.B1007}
  {\bibfield  {journal} {\bibinfo  {journal} {Phys. Rev.}\ }\textbf {\bibinfo
  {volume} {138}},\ \bibinfo {pages} {B1007} (\bibinfo {year}
  {1965})}\BibitemShut {NoStop}%
\bibitem [{\citenamefont {Rivas}\ \emph {et~al.}(2010)\citenamefont {Rivas},
  \citenamefont {Plato}, \citenamefont {Huelga},\ and\ \citenamefont
  {Plenio}}]{rivas2010markovian}%
  \BibitemOpen
  \bibfield  {author} {\bibinfo {author} {\bibfnamefont {A.}~\bibnamefont
  {Rivas}}, \bibinfo {author} {\bibfnamefont {A.~D.~K.}\ \bibnamefont {Plato}},
  \bibinfo {author} {\bibfnamefont {S.~F.}\ \bibnamefont {Huelga}},\ and\
  \bibinfo {author} {\bibfnamefont {M.~B.}\ \bibnamefont {Plenio}},\ }\bibfield
   {title} {\bibinfo {title} {Markovian master equations: a critical study},\
  }\href@noop {} {\bibfield  {journal} {\bibinfo  {journal} {New Journal of
  Physics}\ }\textbf {\bibinfo {volume} {12}},\ \bibinfo {pages} {113032}
  (\bibinfo {year} {2010})}\BibitemShut {NoStop}%
\bibitem [{\citenamefont {Albash}\ \emph {et~al.}(2012)\citenamefont {Albash},
  \citenamefont {Boixo}, \citenamefont {Lidar},\ and\ \citenamefont
  {Zanardi}}]{albash2012quantum}%
  \BibitemOpen
  \bibfield  {author} {\bibinfo {author} {\bibfnamefont {T.}~\bibnamefont
  {Albash}}, \bibinfo {author} {\bibfnamefont {S.}~\bibnamefont {Boixo}},
  \bibinfo {author} {\bibfnamefont {D.~A.}\ \bibnamefont {Lidar}},\ and\
  \bibinfo {author} {\bibfnamefont {P.}~\bibnamefont {Zanardi}},\ }\bibfield
  {title} {\bibinfo {title} {Quantum adiabatic markovian master equations},\
  }\href@noop {} {\bibfield  {journal} {\bibinfo  {journal} {New Journal of
  Physics}\ }\textbf {\bibinfo {volume} {14}},\ \bibinfo {pages} {123016}
  (\bibinfo {year} {2012})}\BibitemShut {NoStop}%
\bibitem [{\citenamefont {Gardiner}\ and\ \citenamefont
  {Zoller}(2004)}]{gardiner2004quantum}%
  \BibitemOpen
  \bibfield  {author} {\bibinfo {author} {\bibfnamefont {C.}~\bibnamefont
  {Gardiner}}\ and\ \bibinfo {author} {\bibfnamefont {P.}~\bibnamefont
  {Zoller}},\ }\href@noop {} {\emph {\bibinfo {title} {Quantum noise: a
  handbook of Markovian and non-Markovian quantum stochastic methods with
  applications to quantum optics}}}\ (\bibinfo  {publisher} {Springer Science
  \& Business Media},\ \bibinfo {year} {2004})\BibitemShut {NoStop}%
\bibitem [{\citenamefont {Mozgunov}\ and\ \citenamefont
  {Lidar}(2020)}]{mozgunov2020completely}%
  \BibitemOpen
  \bibfield  {author} {\bibinfo {author} {\bibfnamefont {E.}~\bibnamefont
  {Mozgunov}}\ and\ \bibinfo {author} {\bibfnamefont {D.}~\bibnamefont
  {Lidar}},\ }\bibfield  {title} {\bibinfo {title} {Completely positive master
  equation for arbitrary driving and small level spacing},\ }\href@noop {}
  {\bibfield  {journal} {\bibinfo  {journal} {Quantum}\ }\textbf {\bibinfo
  {volume} {4}},\ \bibinfo {pages} {227} (\bibinfo {year} {2020})}\BibitemShut
  {NoStop}%
\bibitem [{\citenamefont {Harris}(1996)}]{harris1996encyclopedia}%
  \BibitemOpen
  \bibfield  {author} {\bibinfo {author} {\bibfnamefont {R.}~\bibnamefont
  {Harris}},\ }\bibfield  {title} {\bibinfo {title} {Encyclopedia of nuclear
  magnetic resonance},\ }\href@noop {} {\bibfield  {journal} {\bibinfo
  {journal} {in-Chief DM Grant and RK Harris}\ }\textbf {\bibinfo {volume}
  {5}},\ \bibinfo {pages} {3301} (\bibinfo {year} {1996})}\BibitemShut
  {NoStop}%
\bibitem [{\citenamefont {Ishizaki}\ and\ \citenamefont
  {Fleming}(2009)}]{ishizaki2009adequacy}%
  \BibitemOpen
  \bibfield  {author} {\bibinfo {author} {\bibfnamefont {A.}~\bibnamefont
  {Ishizaki}}\ and\ \bibinfo {author} {\bibfnamefont {G.~R.}\ \bibnamefont
  {Fleming}},\ }\bibfield  {title} {\bibinfo {title} {On the adequacy of the
  redfield equation and related approaches to the study of quantum dynamics in
  electronic energy transfer},\ }\href@noop {} {\bibfield  {journal} {\bibinfo
  {journal} {The Journal of chemical physics}\ }\textbf {\bibinfo {volume}
  {130}} (\bibinfo {year} {2009})}\BibitemShut {NoStop}%
\bibitem [{\citenamefont {Pollard}\ \emph {et~al.}(1996)\citenamefont
  {Pollard}, \citenamefont {Felts},\ and\ \citenamefont
  {Friesner}}]{pollard1996redfield}%
  \BibitemOpen
  \bibfield  {author} {\bibinfo {author} {\bibfnamefont {W.~T.}\ \bibnamefont
  {Pollard}}, \bibinfo {author} {\bibfnamefont {A.~K.}\ \bibnamefont {Felts}},\
  and\ \bibinfo {author} {\bibfnamefont {R.~A.}\ \bibnamefont {Friesner}},\
  }\bibfield  {title} {\bibinfo {title} {The redfield equation in
  condensed-phase quantum dynamics},\ }\href@noop {} {\bibfield  {journal}
  {\bibinfo  {journal} {Advances in Chemical Physics (John Wiley \& Sons, Inc.,
  2007) pp}\ ,\ \bibinfo {pages} {77}} (\bibinfo {year} {1996})}\BibitemShut
  {NoStop}%
\bibitem [{\citenamefont {Gemmer}\ \emph {et~al.}(2009)\citenamefont {Gemmer},
  \citenamefont {Michel},\ and\ \citenamefont {Mahler}}]{gemmer2009quantum}%
  \BibitemOpen
  \bibfield  {author} {\bibinfo {author} {\bibfnamefont {J.}~\bibnamefont
  {Gemmer}}, \bibinfo {author} {\bibfnamefont {M.}~\bibnamefont {Michel}},\
  and\ \bibinfo {author} {\bibfnamefont {G.}~\bibnamefont {Mahler}},\
  }\href@noop {} {\emph {\bibinfo {title} {Quantum thermodynamics: Emergence of
  thermodynamic behavior within composite quantum systems}}},\ Vol.\ \bibinfo
  {volume} {784}\ (\bibinfo  {publisher} {Springer},\ \bibinfo {year}
  {2009})\BibitemShut {NoStop}%
\bibitem [{\citenamefont {Alicki}\ \emph {et~al.}(2006)\citenamefont {Alicki},
  \citenamefont {Lidar},\ and\ \citenamefont {Zanardi}}]{alicki2006internal}%
  \BibitemOpen
  \bibfield  {author} {\bibinfo {author} {\bibfnamefont {R.}~\bibnamefont
  {Alicki}}, \bibinfo {author} {\bibfnamefont {D.~A.}\ \bibnamefont {Lidar}},\
  and\ \bibinfo {author} {\bibfnamefont {P.}~\bibnamefont {Zanardi}},\
  }\bibfield  {title} {\bibinfo {title} {Internal consistency of fault-tolerant
  quantum error correction in light of rigorous derivations of the quantum
  markovian limit},\ }\href@noop {} {\bibfield  {journal} {\bibinfo  {journal}
  {Physical Review A}\ }\textbf {\bibinfo {volume} {73}},\ \bibinfo {pages}
  {052311} (\bibinfo {year} {2006})}\BibitemShut {NoStop}%
\bibitem [{\citenamefont {Ptaszy\ifmmode~\acute{n}\else \'{n}\fi{}ski}\ and\
  \citenamefont {Esposito}(2019)}]{PhysRevLett.122.150603}%
  \BibitemOpen
  \bibfield  {author} {\bibinfo {author} {\bibfnamefont {K.}~\bibnamefont
  {Ptaszy\ifmmode~\acute{n}\else \'{n}\fi{}ski}}\ and\ \bibinfo {author}
  {\bibfnamefont {M.}~\bibnamefont {Esposito}},\ }\bibfield  {title} {\bibinfo
  {title} {Thermodynamics of quantum information flows},\ }\href
  {https://doi.org/10.1103/PhysRevLett.122.150603} {\bibfield  {journal}
  {\bibinfo  {journal} {Phys. Rev. Lett.}\ }\textbf {\bibinfo {volume} {122}},\
  \bibinfo {pages} {150603} (\bibinfo {year} {2019})}\BibitemShut {NoStop}%
\bibitem [{\citenamefont {Wang}\ \emph {et~al.}(2023)\citenamefont {Wang},
  \citenamefont {Yang},\ and\ \citenamefont {Zhang}}]{PhysRevE.107.014108}%
  \BibitemOpen
  \bibfield  {author} {\bibinfo {author} {\bibfnamefont {S.-Y.}\ \bibnamefont
  {Wang}}, \bibinfo {author} {\bibfnamefont {Q.}~\bibnamefont {Yang}},\ and\
  \bibinfo {author} {\bibfnamefont {F.-L.}\ \bibnamefont {Zhang}},\ }\bibfield
  {title} {\bibinfo {title} {Thermodynamically consistent master equation based
  on subsystem eigenstates},\ }\href
  {https://doi.org/10.1103/PhysRevE.107.014108} {\bibfield  {journal} {\bibinfo
   {journal} {Phys. Rev. E}\ }\textbf {\bibinfo {volume} {107}},\ \bibinfo
  {pages} {014108} (\bibinfo {year} {2023})}\BibitemShut {NoStop}%
\bibitem [{\citenamefont {REDFIELD}(1965)}]{REDFIELD19651}%
  \BibitemOpen
  \bibfield  {author} {\bibinfo {author} {\bibfnamefont {A.}~\bibnamefont
  {REDFIELD}},\ }\bibfield  {title} {\bibinfo {title} {The theory of relaxation
  processes},\ }in\ \href
  {https://doi.org/https://doi.org/10.1016/B978-1-4832-3114-3.50007-6} {\emph
  {\bibinfo {booktitle} {Advances in Magnetic Resonance}}},\ \bibinfo {series}
  {Advances in Magnetic and Optical Resonance}, Vol.~\bibinfo {volume} {1},\
  \bibinfo {editor} {edited by\ \bibinfo {editor} {\bibfnamefont {J.~S.}\
  \bibnamefont {Waugh}}}\ (\bibinfo  {publisher} {Academic Press},\ \bibinfo
  {year} {1965})\ pp.\ \bibinfo {pages} {1--32}\BibitemShut {NoStop}%
\bibitem [{\citenamefont {Redfield}(1957)}]{redfield1957theory}%
  \BibitemOpen
  \bibfield  {author} {\bibinfo {author} {\bibfnamefont {A.~G.}\ \bibnamefont
  {Redfield}},\ }\bibfield  {title} {\bibinfo {title} {On the theory of
  relaxation processes},\ }\href@noop {} {\bibfield  {journal} {\bibinfo
  {journal} {IBM Journal of Research and Development}\ }\textbf {\bibinfo
  {volume} {1}},\ \bibinfo {pages} {19} (\bibinfo {year} {1957})}\BibitemShut
  {NoStop}%
\bibitem [{\citenamefont {Lindblad}(1976)}]{lindblad1976generators}%
  \BibitemOpen
  \bibfield  {author} {\bibinfo {author} {\bibfnamefont {G.}~\bibnamefont
  {Lindblad}},\ }\bibfield  {title} {\bibinfo {title} {On the generators of
  quantum dynamical semigroups},\ }\href@noop {} {\bibfield  {journal}
  {\bibinfo  {journal} {Communications in mathematical physics}\ }\textbf
  {\bibinfo {volume} {48}},\ \bibinfo {pages} {119} (\bibinfo {year}
  {1976})}\BibitemShut {NoStop}%
\bibitem [{\citenamefont {Davies}(1974)}]{davies1974markovian}%
  \BibitemOpen
  \bibfield  {author} {\bibinfo {author} {\bibfnamefont {E.~B.}\ \bibnamefont
  {Davies}},\ }\bibfield  {title} {\bibinfo {title} {Markovian master
  equations},\ }\href@noop {} {\bibfield  {journal} {\bibinfo  {journal}
  {Communications in mathematical Physics}\ }\textbf {\bibinfo {volume} {39}},\
  \bibinfo {pages} {91} (\bibinfo {year} {1974})}\BibitemShut {NoStop}%
\bibitem [{\citenamefont {Gorini}\ \emph {et~al.}(1976)\citenamefont {Gorini},
  \citenamefont {Kossakowski},\ and\ \citenamefont
  {Sudarshan}}]{gorini1976completely}%
  \BibitemOpen
  \bibfield  {author} {\bibinfo {author} {\bibfnamefont {V.}~\bibnamefont
  {Gorini}}, \bibinfo {author} {\bibfnamefont {A.}~\bibnamefont
  {Kossakowski}},\ and\ \bibinfo {author} {\bibfnamefont {E.~C.~G.}\
  \bibnamefont {Sudarshan}},\ }\bibfield  {title} {\bibinfo {title} {Completely
  positive dynamical semigroups of n-level systems},\ }\href@noop {} {\bibfield
   {journal} {\bibinfo  {journal} {Journal of Mathematical Physics}\ }\textbf
  {\bibinfo {volume} {17}},\ \bibinfo {pages} {821} (\bibinfo {year}
  {1976})}\BibitemShut {NoStop}%
\bibitem [{\citenamefont {Vadimov}\ \emph {et~al.}(2021)\citenamefont
  {Vadimov}, \citenamefont {Tuorila}, \citenamefont {Orell}, \citenamefont
  {Stockburger}, \citenamefont {Ala-Nissila}, \citenamefont {Ankerhold},\ and\
  \citenamefont {M\"ott\"onen}}]{PhysRevB.103.214308}%
  \BibitemOpen
  \bibfield  {author} {\bibinfo {author} {\bibfnamefont {V.}~\bibnamefont
  {Vadimov}}, \bibinfo {author} {\bibfnamefont {J.}~\bibnamefont {Tuorila}},
  \bibinfo {author} {\bibfnamefont {T.}~\bibnamefont {Orell}}, \bibinfo
  {author} {\bibfnamefont {J.}~\bibnamefont {Stockburger}}, \bibinfo {author}
  {\bibfnamefont {T.}~\bibnamefont {Ala-Nissila}}, \bibinfo {author}
  {\bibfnamefont {J.}~\bibnamefont {Ankerhold}},\ and\ \bibinfo {author}
  {\bibfnamefont {M.}~\bibnamefont {M\"ott\"onen}},\ }\bibfield  {title}
  {\bibinfo {title} {Validity of born-markov master equations for single- and
  two-qubit systems},\ }\href {https://doi.org/10.1103/PhysRevB.103.214308}
  {\bibfield  {journal} {\bibinfo  {journal} {Phys. Rev. B}\ }\textbf {\bibinfo
  {volume} {103}},\ \bibinfo {pages} {214308} (\bibinfo {year}
  {2021})}\BibitemShut {NoStop}%
\bibitem [{\citenamefont {Davidovi{\'c}}(2020)}]{davidovic2020completely}%
  \BibitemOpen
  \bibfield  {author} {\bibinfo {author} {\bibfnamefont {D.}~\bibnamefont
  {Davidovi{\'c}}},\ }\bibfield  {title} {\bibinfo {title} {Completely
  positive, simple, and possibly highly accurate approximation of the redfield
  equation},\ }\href@noop {} {\bibfield  {journal} {\bibinfo  {journal}
  {Quantum}\ }\textbf {\bibinfo {volume} {4}},\ \bibinfo {pages} {326}
  (\bibinfo {year} {2020})}\BibitemShut {NoStop}%
\bibitem [{\citenamefont {Lee}\ and\ \citenamefont
  {Yeo}(2022)}]{PhysRevE.106.054145}%
  \BibitemOpen
  \bibfield  {author} {\bibinfo {author} {\bibfnamefont {J.~S.}\ \bibnamefont
  {Lee}}\ and\ \bibinfo {author} {\bibfnamefont {J.}~\bibnamefont {Yeo}},\
  }\bibfield  {title} {\bibinfo {title} {Perturbative steady states of
  completely positive quantum master equations},\ }\href
  {https://doi.org/10.1103/PhysRevE.106.054145} {\bibfield  {journal} {\bibinfo
   {journal} {Phys. Rev. E}\ }\textbf {\bibinfo {volume} {106}},\ \bibinfo
  {pages} {054145} (\bibinfo {year} {2022})}\BibitemShut {NoStop}%
\bibitem [{\citenamefont {Trushechkin}\ \emph {et~al.}(2022)\citenamefont
  {Trushechkin}, \citenamefont {Merkli}, \citenamefont {Cresser},\ and\
  \citenamefont {Anders}}]{10.1116/5.0073853}%
  \BibitemOpen
  \bibfield  {author} {\bibinfo {author} {\bibfnamefont {A.~S.}\ \bibnamefont
  {Trushechkin}}, \bibinfo {author} {\bibfnamefont {M.}~\bibnamefont {Merkli}},
  \bibinfo {author} {\bibfnamefont {J.~D.}\ \bibnamefont {Cresser}},\ and\
  \bibinfo {author} {\bibfnamefont {J.}~\bibnamefont {Anders}},\ }\bibfield
  {title} {\bibinfo {title} {Open quantum system dynamics and the mean force
  gibbs state},\ }\href {https://doi.org/10.1116/5.0073853} {\bibfield
  {journal} {\bibinfo  {journal} {AVS Quantum Science}\ }\textbf {\bibinfo
  {volume} {4}},\ \bibinfo {pages} {012301} (\bibinfo {year} {2022})},\ \Eprint
  {https://arxiv.org/abs/https://pubs.aip.org/avs/aqs/article-pdf/doi/10.1116/5.0073853/19803511/012301\_1\_online.pdf}
  {https://pubs.aip.org/avs/aqs/article-pdf/doi/10.1116/5.0073853/19803511/012301\_1\_online.pdf}
  \BibitemShut {NoStop}%
\bibitem [{\citenamefont {Cresser}\ and\ \citenamefont
  {Anders}(2021)}]{PhysRevLett.127.250601}%
  \BibitemOpen
  \bibfield  {author} {\bibinfo {author} {\bibfnamefont {J.~D.}\ \bibnamefont
  {Cresser}}\ and\ \bibinfo {author} {\bibfnamefont {J.}~\bibnamefont
  {Anders}},\ }\bibfield  {title} {\bibinfo {title} {Weak and ultrastrong
  coupling limits of the quantum mean force gibbs state},\ }\href
  {https://doi.org/10.1103/PhysRevLett.127.250601} {\bibfield  {journal}
  {\bibinfo  {journal} {Phys. Rev. Lett.}\ }\textbf {\bibinfo {volume} {127}},\
  \bibinfo {pages} {250601} (\bibinfo {year} {2021})}\BibitemShut {NoStop}%
\bibitem [{\citenamefont {Thingna}\ \emph {et~al.}(2012)\citenamefont
  {Thingna}, \citenamefont {Wang},\ and\ \citenamefont
  {Hänggi}}]{10.1063/1.4718706}%
  \BibitemOpen
  \bibfield  {author} {\bibinfo {author} {\bibfnamefont {J.}~\bibnamefont
  {Thingna}}, \bibinfo {author} {\bibfnamefont {J.-S.}\ \bibnamefont {Wang}},\
  and\ \bibinfo {author} {\bibfnamefont {P.}~\bibnamefont {Hänggi}},\
  }\bibfield  {title} {\bibinfo {title} {Generalized gibbs state with modified
  redfield solution: Exact agreement up to second order},\ }\href
  {https://doi.org/10.1063/1.4718706} {\bibfield  {journal} {\bibinfo
  {journal} {The Journal of Chemical Physics}\ }\textbf {\bibinfo {volume}
  {136}},\ \bibinfo {pages} {194110} (\bibinfo {year} {2012})},\ \Eprint
  {https://arxiv.org/abs/https://pubs.aip.org/aip/jcp/article-pdf/doi/10.1063/1.4718706/15449334/194110\_1\_online.pdf}
  {https://pubs.aip.org/aip/jcp/article-pdf/doi/10.1063/1.4718706/15449334/194110\_1\_online.pdf}
  \BibitemShut {NoStop}%
\bibitem [{\citenamefont {Segal}(2006)}]{PhysRevB.73.205415}%
  \BibitemOpen
  \bibfield  {author} {\bibinfo {author} {\bibfnamefont {D.}~\bibnamefont
  {Segal}},\ }\bibfield  {title} {\bibinfo {title} {Heat flow in nonlinear
  molecular junctions: Master equation analysis},\ }\href
  {https://doi.org/10.1103/PhysRevB.73.205415} {\bibfield  {journal} {\bibinfo
  {journal} {Phys. Rev. B}\ }\textbf {\bibinfo {volume} {73}},\ \bibinfo
  {pages} {205415} (\bibinfo {year} {2006})}\BibitemShut {NoStop}%
\bibitem [{\citenamefont {Cao}\ \emph {et~al.}(2021)\citenamefont {Cao},
  \citenamefont {Wang}, \citenamefont {Zheng},\ and\ \citenamefont
  {He}}]{PhysRevB.103.075407}%
  \BibitemOpen
  \bibfield  {author} {\bibinfo {author} {\bibfnamefont {X.}~\bibnamefont
  {Cao}}, \bibinfo {author} {\bibfnamefont {C.}~\bibnamefont {Wang}}, \bibinfo
  {author} {\bibfnamefont {H.}~\bibnamefont {Zheng}},\ and\ \bibinfo {author}
  {\bibfnamefont {D.}~\bibnamefont {He}},\ }\bibfield  {title} {\bibinfo
  {title} {Quantum thermal transport via a canonically transformed redfield
  approach},\ }\href {https://doi.org/10.1103/PhysRevB.103.075407} {\bibfield
  {journal} {\bibinfo  {journal} {Phys. Rev. B}\ }\textbf {\bibinfo {volume}
  {103}},\ \bibinfo {pages} {075407} (\bibinfo {year} {2021})}\BibitemShut
  {NoStop}%
\bibitem [{\citenamefont {Dey}\ \emph {et~al.}(2023)\citenamefont {Dey},
  \citenamefont {Timossi}, \citenamefont {Amico},\ and\ \citenamefont
  {Marchegiani}}]{PhysRevB.107.134510}%
  \BibitemOpen
  \bibfield  {author} {\bibinfo {author} {\bibfnamefont {S.~S.}\ \bibnamefont
  {Dey}}, \bibinfo {author} {\bibfnamefont {G.}~\bibnamefont {Timossi}},
  \bibinfo {author} {\bibfnamefont {L.}~\bibnamefont {Amico}},\ and\ \bibinfo
  {author} {\bibfnamefont {G.}~\bibnamefont {Marchegiani}},\ }\bibfield
  {title} {\bibinfo {title} {Negative differential thermal conductance by
  photonic transport in electronic circuits},\ }\href
  {https://doi.org/10.1103/PhysRevB.107.134510} {\bibfield  {journal} {\bibinfo
   {journal} {Phys. Rev. B}\ }\textbf {\bibinfo {volume} {107}},\ \bibinfo
  {pages} {134510} (\bibinfo {year} {2023})}\BibitemShut {NoStop}%
\bibitem [{\citenamefont {Liu}\ \emph {et~al.}(2019)\citenamefont {Liu},
  \citenamefont {Wang}, \citenamefont {Wang},\ and\ \citenamefont
  {Ren}}]{PhysRevE.99.032114}%
  \BibitemOpen
  \bibfield  {author} {\bibinfo {author} {\bibfnamefont {H.}~\bibnamefont
  {Liu}}, \bibinfo {author} {\bibfnamefont {C.}~\bibnamefont {Wang}}, \bibinfo
  {author} {\bibfnamefont {L.-Q.}\ \bibnamefont {Wang}},\ and\ \bibinfo
  {author} {\bibfnamefont {J.}~\bibnamefont {Ren}},\ }\bibfield  {title}
  {\bibinfo {title} {Strong system-bath coupling induces negative differential
  thermal conductance and heat amplification in nonequilibrium two-qubit
  systems},\ }\href {https://doi.org/10.1103/PhysRevE.99.032114} {\bibfield
  {journal} {\bibinfo  {journal} {Phys. Rev. E}\ }\textbf {\bibinfo {volume}
  {99}},\ \bibinfo {pages} {032114} (\bibinfo {year} {2019})}\BibitemShut
  {NoStop}%
\bibitem [{\citenamefont {Joulain}\ \emph {et~al.}(2015)\citenamefont
  {Joulain}, \citenamefont {Ezzahri}, \citenamefont {Drevillon},\ and\
  \citenamefont {Ben-Abdallah}}]{joulain2015modulation}%
  \BibitemOpen
  \bibfield  {author} {\bibinfo {author} {\bibfnamefont {K.}~\bibnamefont
  {Joulain}}, \bibinfo {author} {\bibfnamefont {Y.}~\bibnamefont {Ezzahri}},
  \bibinfo {author} {\bibfnamefont {J.}~\bibnamefont {Drevillon}},\ and\
  \bibinfo {author} {\bibfnamefont {P.}~\bibnamefont {Ben-Abdallah}},\
  }\bibfield  {title} {\bibinfo {title} {Modulation and amplification of
  radiative far field heat transfer: Towards a simple radiative thermal
  transistor},\ }\href@noop {} {\bibfield  {journal} {\bibinfo  {journal}
  {Applied Physics Letters}\ }\textbf {\bibinfo {volume} {106}},\ \bibinfo
  {pages} {133505} (\bibinfo {year} {2015})}\BibitemShut {NoStop}%
\bibitem [{\citenamefont {Li}\ \emph {et~al.}(2012)\citenamefont {Li},
  \citenamefont {Ren}, \citenamefont {Wang}, \citenamefont {Zhang},
  \citenamefont {H\"anggi},\ and\ \citenamefont {Li}}]{RevModPhys.84.1045}%
  \BibitemOpen
  \bibfield  {author} {\bibinfo {author} {\bibfnamefont {N.}~\bibnamefont
  {Li}}, \bibinfo {author} {\bibfnamefont {J.}~\bibnamefont {Ren}}, \bibinfo
  {author} {\bibfnamefont {L.}~\bibnamefont {Wang}}, \bibinfo {author}
  {\bibfnamefont {G.}~\bibnamefont {Zhang}}, \bibinfo {author} {\bibfnamefont
  {P.}~\bibnamefont {H\"anggi}},\ and\ \bibinfo {author} {\bibfnamefont
  {B.}~\bibnamefont {Li}},\ }\bibfield  {title} {\bibinfo {title} {Colloquium:
  Phononics: Manipulating heat flow with electronic analogs and beyond},\
  }\href {https://doi.org/10.1103/RevModPhys.84.1045} {\bibfield  {journal}
  {\bibinfo  {journal} {Rev. Mod. Phys.}\ }\textbf {\bibinfo {volume} {84}},\
  \bibinfo {pages} {1045} (\bibinfo {year} {2012})}\BibitemShut {NoStop}%
\bibitem [{\citenamefont {Ordonez-Miranda}\ \emph {et~al.}(2017)\citenamefont
  {Ordonez-Miranda}, \citenamefont {Ezzahri},\ and\ \citenamefont
  {Joulain}}]{PhysRevE.95.022128}%
  \BibitemOpen
  \bibfield  {author} {\bibinfo {author} {\bibfnamefont {J.}~\bibnamefont
  {Ordonez-Miranda}}, \bibinfo {author} {\bibfnamefont {Y.}~\bibnamefont
  {Ezzahri}},\ and\ \bibinfo {author} {\bibfnamefont {K.}~\bibnamefont
  {Joulain}},\ }\bibfield  {title} {\bibinfo {title} {Quantum thermal diode
  based on two interacting spinlike systems under different excitations},\
  }\href {https://doi.org/10.1103/PhysRevE.95.022128} {\bibfield  {journal}
  {\bibinfo  {journal} {Phys. Rev. E}\ }\textbf {\bibinfo {volume} {95}},\
  \bibinfo {pages} {022128} (\bibinfo {year} {2017})}\BibitemShut {NoStop}%
\bibitem [{\citenamefont {Karg\ifmmode \imath \else~\i \fi{}}\ \emph
  {et~al.}(2019)\citenamefont {Karg\ifmmode \imath \else~\i \fi{}},
  \citenamefont {Naseem}, \citenamefont {Opatrn\'y}, \citenamefont
  {M\"ustecapl\ifmmode \imath \else \i \fi{}o\ifmmode~\breve{g}\else
  \u{g}\fi{}lu},\ and\ \citenamefont {Kurizki}}]{PhysRevE.99.042121}%
  \BibitemOpen
  \bibfield  {author} {\bibinfo {author} {\bibfnamefont {C.}~\bibnamefont
  {Karg\ifmmode \imath \else~\i \fi{}}}, \bibinfo {author} {\bibfnamefont
  {M.~T.}\ \bibnamefont {Naseem}}, \bibinfo {author} {\bibfnamefont {T.~c.~v.}\
  \bibnamefont {Opatrn\'y}}, \bibinfo {author} {\bibfnamefont {O.~E.}\
  \bibnamefont {M\"ustecapl\ifmmode \imath \else \i
  \fi{}o\ifmmode~\breve{g}\else \u{g}\fi{}lu}},\ and\ \bibinfo {author}
  {\bibfnamefont {G.}~\bibnamefont {Kurizki}},\ }\bibfield  {title} {\bibinfo
  {title} {Quantum optical two-atom thermal diode},\ }\href
  {https://doi.org/10.1103/PhysRevE.99.042121} {\bibfield  {journal} {\bibinfo
  {journal} {Phys. Rev. E}\ }\textbf {\bibinfo {volume} {99}},\ \bibinfo
  {pages} {042121} (\bibinfo {year} {2019})}\BibitemShut {NoStop}%
\bibitem [{\citenamefont {Wang}\ \emph
  {et~al.}(2019{\natexlab{a}})\citenamefont {Wang}, \citenamefont {Xu},
  \citenamefont {Liu},\ and\ \citenamefont {Gao}}]{PhysRevE.99.042102}%
  \BibitemOpen
  \bibfield  {author} {\bibinfo {author} {\bibfnamefont {C.}~\bibnamefont
  {Wang}}, \bibinfo {author} {\bibfnamefont {D.}~\bibnamefont {Xu}}, \bibinfo
  {author} {\bibfnamefont {H.}~\bibnamefont {Liu}},\ and\ \bibinfo {author}
  {\bibfnamefont {X.}~\bibnamefont {Gao}},\ }\bibfield  {title} {\bibinfo
  {title} {Thermal rectification and heat amplification in a nonequilibrium
  v-type three-level system},\ }\href
  {https://doi.org/10.1103/PhysRevE.99.042102} {\bibfield  {journal} {\bibinfo
  {journal} {Phys. Rev. E}\ }\textbf {\bibinfo {volume} {99}},\ \bibinfo
  {pages} {042102} (\bibinfo {year} {2019}{\natexlab{a}})}\BibitemShut
  {NoStop}%
\bibitem [{\citenamefont {Tesser}\ \emph {et~al.}(2022)\citenamefont {Tesser},
  \citenamefont {Bhandari}, \citenamefont {Erdman}, \citenamefont {Paladino},
  \citenamefont {Fazio},\ and\ \citenamefont {Taddei}}]{Tesser_2022}%
  \BibitemOpen
  \bibfield  {author} {\bibinfo {author} {\bibfnamefont {L.}~\bibnamefont
  {Tesser}}, \bibinfo {author} {\bibfnamefont {B.}~\bibnamefont {Bhandari}},
  \bibinfo {author} {\bibfnamefont {P.~A.}\ \bibnamefont {Erdman}}, \bibinfo
  {author} {\bibfnamefont {E.}~\bibnamefont {Paladino}}, \bibinfo {author}
  {\bibfnamefont {R.}~\bibnamefont {Fazio}},\ and\ \bibinfo {author}
  {\bibfnamefont {F.}~\bibnamefont {Taddei}},\ }\bibfield  {title} {\bibinfo
  {title} {Heat rectification through single and coupled quantum dots},\ }\href
  {https://doi.org/10.1088/1367-2630/ac53b8} {\bibfield  {journal} {\bibinfo
  {journal} {New Journal of Physics}\ }\textbf {\bibinfo {volume} {24}},\
  \bibinfo {pages} {035001} (\bibinfo {year} {2022})}\BibitemShut {NoStop}%
\bibitem [{\citenamefont {Balachandran}\ \emph {et~al.}(2018)\citenamefont
  {Balachandran}, \citenamefont {Benenti}, \citenamefont {Pereira},
  \citenamefont {Casati},\ and\ \citenamefont
  {Poletti}}]{PhysRevLett.120.200603}%
  \BibitemOpen
  \bibfield  {author} {\bibinfo {author} {\bibfnamefont {V.}~\bibnamefont
  {Balachandran}}, \bibinfo {author} {\bibfnamefont {G.}~\bibnamefont
  {Benenti}}, \bibinfo {author} {\bibfnamefont {E.}~\bibnamefont {Pereira}},
  \bibinfo {author} {\bibfnamefont {G.}~\bibnamefont {Casati}},\ and\ \bibinfo
  {author} {\bibfnamefont {D.}~\bibnamefont {Poletti}},\ }\bibfield  {title}
  {\bibinfo {title} {Perfect diode in quantum spin chains},\ }\href
  {https://doi.org/10.1103/PhysRevLett.120.200603} {\bibfield  {journal}
  {\bibinfo  {journal} {Phys. Rev. Lett.}\ }\textbf {\bibinfo {volume} {120}},\
  \bibinfo {pages} {200603} (\bibinfo {year} {2018})}\BibitemShut {NoStop}%
\bibitem [{\citenamefont {Segal}(2008)}]{PhysRevLett.100.105901}%
  \BibitemOpen
  \bibfield  {author} {\bibinfo {author} {\bibfnamefont {D.}~\bibnamefont
  {Segal}},\ }\bibfield  {title} {\bibinfo {title} {Single mode heat rectifier:
  Controlling energy flow between electronic conductors},\ }\href
  {https://doi.org/10.1103/PhysRevLett.100.105901} {\bibfield  {journal}
  {\bibinfo  {journal} {Phys. Rev. Lett.}\ }\textbf {\bibinfo {volume} {100}},\
  \bibinfo {pages} {105901} (\bibinfo {year} {2008})}\BibitemShut {NoStop}%
\bibitem [{\citenamefont {Terraneo}\ \emph {et~al.}(2002)\citenamefont
  {Terraneo}, \citenamefont {Peyrard},\ and\ \citenamefont
  {Casati}}]{PhysRevLett.88.094302}%
  \BibitemOpen
  \bibfield  {author} {\bibinfo {author} {\bibfnamefont {M.}~\bibnamefont
  {Terraneo}}, \bibinfo {author} {\bibfnamefont {M.}~\bibnamefont {Peyrard}},\
  and\ \bibinfo {author} {\bibfnamefont {G.}~\bibnamefont {Casati}},\
  }\bibfield  {title} {\bibinfo {title} {Controlling the energy flow in
  nonlinear lattices: A model for a thermal rectifier},\ }\href
  {https://doi.org/10.1103/PhysRevLett.88.094302} {\bibfield  {journal}
  {\bibinfo  {journal} {Phys. Rev. Lett.}\ }\textbf {\bibinfo {volume} {88}},\
  \bibinfo {pages} {094302} (\bibinfo {year} {2002})}\BibitemShut {NoStop}%
\bibitem [{\citenamefont {Pereira}(2011)}]{PhysRevE.83.031106}%
  \BibitemOpen
  \bibfield  {author} {\bibinfo {author} {\bibfnamefont {E.}~\bibnamefont
  {Pereira}},\ }\bibfield  {title} {\bibinfo {title} {Sufficient conditions for
  thermal rectification in general graded materials},\ }\href
  {https://doi.org/10.1103/PhysRevE.83.031106} {\bibfield  {journal} {\bibinfo
  {journal} {Phys. Rev. E}\ }\textbf {\bibinfo {volume} {83}},\ \bibinfo
  {pages} {031106} (\bibinfo {year} {2011})}\BibitemShut {NoStop}%
\bibitem [{\citenamefont {Yang}\ \emph {et~al.}(2019)\citenamefont {Yang},
  \citenamefont {Elouard}, \citenamefont {Splettstoesser}, \citenamefont
  {Sothmann}, \citenamefont {S\'anchez},\ and\ \citenamefont
  {Jordan}}]{PhysRevB.100.045418}%
  \BibitemOpen
  \bibfield  {author} {\bibinfo {author} {\bibfnamefont {J.}~\bibnamefont
  {Yang}}, \bibinfo {author} {\bibfnamefont {C.}~\bibnamefont {Elouard}},
  \bibinfo {author} {\bibfnamefont {J.}~\bibnamefont {Splettstoesser}},
  \bibinfo {author} {\bibfnamefont {B.}~\bibnamefont {Sothmann}}, \bibinfo
  {author} {\bibfnamefont {R.}~\bibnamefont {S\'anchez}},\ and\ \bibinfo
  {author} {\bibfnamefont {A.~N.}\ \bibnamefont {Jordan}},\ }\bibfield  {title}
  {\bibinfo {title} {Thermal transistor and thermometer based on
  coulomb-coupled conductors},\ }\href
  {https://doi.org/10.1103/PhysRevB.100.045418} {\bibfield  {journal} {\bibinfo
   {journal} {Phys. Rev. B}\ }\textbf {\bibinfo {volume} {100}},\ \bibinfo
  {pages} {045418} (\bibinfo {year} {2019})}\BibitemShut {NoStop}%
\bibitem [{\citenamefont {Hovhannisyan}\ and\ \citenamefont
  {Imparato}(2019)}]{hovhannisyan2019quantum}%
  \BibitemOpen
  \bibfield  {author} {\bibinfo {author} {\bibfnamefont {K.~V.}\ \bibnamefont
  {Hovhannisyan}}\ and\ \bibinfo {author} {\bibfnamefont {A.}~\bibnamefont
  {Imparato}},\ }\bibfield  {title} {\bibinfo {title} {Quantum current in
  dissipative systems},\ }\href@noop {} {\bibfield  {journal} {\bibinfo
  {journal} {New Journal of Physics}\ }\textbf {\bibinfo {volume} {21}},\
  \bibinfo {pages} {052001} (\bibinfo {year} {2019})}\BibitemShut {NoStop}%
\bibitem [{\citenamefont {Quiroga}\ \emph {et~al.}(2007)\citenamefont
  {Quiroga}, \citenamefont {Rodr\'{\i}guez}, \citenamefont {Ram\'{\i}rez},\
  and\ \citenamefont {Par\'{\i}s}}]{PhysRevA.75.032308}%
  \BibitemOpen
  \bibfield  {author} {\bibinfo {author} {\bibfnamefont {L.}~\bibnamefont
  {Quiroga}}, \bibinfo {author} {\bibfnamefont {F.~J.}\ \bibnamefont
  {Rodr\'{\i}guez}}, \bibinfo {author} {\bibfnamefont {M.~E.}\ \bibnamefont
  {Ram\'{\i}rez}},\ and\ \bibinfo {author} {\bibfnamefont {R.}~\bibnamefont
  {Par\'{\i}s}},\ }\bibfield  {title} {\bibinfo {title} {Nonequilibrium thermal
  entanglement},\ }\href {https://doi.org/10.1103/PhysRevA.75.032308}
  {\bibfield  {journal} {\bibinfo  {journal} {Phys. Rev. A}\ }\textbf {\bibinfo
  {volume} {75}},\ \bibinfo {pages} {032308} (\bibinfo {year}
  {2007})}\BibitemShut {NoStop}%
\bibitem [{\citenamefont {Sinaysky}\ \emph {et~al.}(2008)\citenamefont
  {Sinaysky}, \citenamefont {Petruccione},\ and\ \citenamefont
  {Burgarth}}]{PhysRevA.78.062301}%
  \BibitemOpen
  \bibfield  {author} {\bibinfo {author} {\bibfnamefont {I.}~\bibnamefont
  {Sinaysky}}, \bibinfo {author} {\bibfnamefont {F.}~\bibnamefont
  {Petruccione}},\ and\ \bibinfo {author} {\bibfnamefont {D.}~\bibnamefont
  {Burgarth}},\ }\bibfield  {title} {\bibinfo {title} {Dynamics of
  nonequilibrium thermal entanglement},\ }\href
  {https://doi.org/10.1103/PhysRevA.78.062301} {\bibfield  {journal} {\bibinfo
  {journal} {Phys. Rev. A}\ }\textbf {\bibinfo {volume} {78}},\ \bibinfo
  {pages} {062301} (\bibinfo {year} {2008})}\BibitemShut {NoStop}%
\bibitem [{\citenamefont {Yang}\ \emph {et~al.}(2022)\citenamefont {Yang},
  \citenamefont {Liu},\ and\ \citenamefont {Yu}}]{yang2022heat}%
  \BibitemOpen
  \bibfield  {author} {\bibinfo {author} {\bibfnamefont {Y.-j.}\ \bibnamefont
  {Yang}}, \bibinfo {author} {\bibfnamefont {Y.-q.}\ \bibnamefont {Liu}},\ and\
  \bibinfo {author} {\bibfnamefont {C.-s.}\ \bibnamefont {Yu}},\ }\bibfield
  {title} {\bibinfo {title} {Heat transfer in transversely coupled qubits:
  optically controlled thermal modulator with common reservoirs},\ }\href@noop
  {} {\bibfield  {journal} {\bibinfo  {journal} {Journal of Physics A:
  Mathematical and Theoretical}\ }\textbf {\bibinfo {volume} {55}},\ \bibinfo
  {pages} {395303} (\bibinfo {year} {2022})}\BibitemShut {NoStop}%
\bibitem [{\citenamefont {Hu}\ \emph {et~al.}(2018)\citenamefont {Hu},
  \citenamefont {Man},\ and\ \citenamefont {Xia}}]{hu2018steady}%
  \BibitemOpen
  \bibfield  {author} {\bibinfo {author} {\bibfnamefont {L.-Z.}\ \bibnamefont
  {Hu}}, \bibinfo {author} {\bibfnamefont {Z.-X.}\ \bibnamefont {Man}},\ and\
  \bibinfo {author} {\bibfnamefont {Y.-J.}\ \bibnamefont {Xia}},\ }\bibfield
  {title} {\bibinfo {title} {Steady-state entanglement and thermalization of
  coupled qubits in two common heat baths},\ }\href@noop {} {\bibfield
  {journal} {\bibinfo  {journal} {Quantum Information Processing}\ }\textbf
  {\bibinfo {volume} {17}},\ \bibinfo {pages} {1} (\bibinfo {year}
  {2018})}\BibitemShut {NoStop}%
\bibitem [{\citenamefont {Cattaneo}\ \emph {et~al.}(2019)\citenamefont
  {Cattaneo}, \citenamefont {Giorgi}, \citenamefont {Maniscalco},\ and\
  \citenamefont {Zambrini}}]{cattaneo2019local}%
  \BibitemOpen
  \bibfield  {author} {\bibinfo {author} {\bibfnamefont {M.}~\bibnamefont
  {Cattaneo}}, \bibinfo {author} {\bibfnamefont {G.~L.}\ \bibnamefont
  {Giorgi}}, \bibinfo {author} {\bibfnamefont {S.}~\bibnamefont {Maniscalco}},\
  and\ \bibinfo {author} {\bibfnamefont {R.}~\bibnamefont {Zambrini}},\
  }\bibfield  {title} {\bibinfo {title} {Local versus global master equation
  with common and separate baths: superiority of the global approach in partial
  secular approximation},\ }\href@noop {} {\bibfield  {journal} {\bibinfo
  {journal} {New Journal of Physics}\ }\textbf {\bibinfo {volume} {21}},\
  \bibinfo {pages} {113045} (\bibinfo {year} {2019})}\BibitemShut {NoStop}%
\bibitem [{\citenamefont {Wang}\ \emph
  {et~al.}(2019{\natexlab{b}})\citenamefont {Wang}, \citenamefont {Wu},\ and\
  \citenamefont {Wang}}]{PhysRevA.99.042320}%
  \BibitemOpen
  \bibfield  {author} {\bibinfo {author} {\bibfnamefont {Z.}~\bibnamefont
  {Wang}}, \bibinfo {author} {\bibfnamefont {W.}~\bibnamefont {Wu}},\ and\
  \bibinfo {author} {\bibfnamefont {J.}~\bibnamefont {Wang}},\ }\bibfield
  {title} {\bibinfo {title} {Steady-state entanglement and coherence of two
  coupled qubits in equilibrium and nonequilibrium environments},\ }\href
  {https://doi.org/10.1103/PhysRevA.99.042320} {\bibfield  {journal} {\bibinfo
  {journal} {Phys. Rev. A}\ }\textbf {\bibinfo {volume} {99}},\ \bibinfo
  {pages} {042320} (\bibinfo {year} {2019}{\natexlab{b}})}\BibitemShut
  {NoStop}%
\bibitem [{\citenamefont {Novotný}(2002)}]{Novotný_2002}%
  \BibitemOpen
  \bibfield  {author} {\bibinfo {author} {\bibfnamefont {T.}~\bibnamefont
  {Novotný}},\ }\bibfield  {title} {\bibinfo {title} {Investigation of
  apparent violation of the second law of thermodynamics in quantum transport
  studies},\ }\href {https://doi.org/10.1209/epl/i2002-00174-3} {\bibfield
  {journal} {\bibinfo  {journal} {Europhysics Letters}\ }\textbf {\bibinfo
  {volume} {59}},\ \bibinfo {pages} {648} (\bibinfo {year} {2002})}\BibitemShut
  {NoStop}%
\bibitem [{\citenamefont {Levy}\ and\ \citenamefont
  {Kosloff}(2014)}]{Levy_2014}%
  \BibitemOpen
  \bibfield  {author} {\bibinfo {author} {\bibfnamefont {A.}~\bibnamefont
  {Levy}}\ and\ \bibinfo {author} {\bibfnamefont {R.}~\bibnamefont {Kosloff}},\
  }\bibfield  {title} {\bibinfo {title} {The local approach to quantum
  transport may violate the second law of thermodynamics},\ }\href
  {https://doi.org/10.1209/0295-5075/107/20004} {\bibfield  {journal} {\bibinfo
   {journal} {Europhysics Letters}\ }\textbf {\bibinfo {volume} {107}},\
  \bibinfo {pages} {20004} (\bibinfo {year} {2014})}\BibitemShut {NoStop}%
\end{thebibliography}%
\appendix
\begin{widetext}
\section{The Derivation of the master equation}
\label{appendixA}
In the interaction picture, the von Neumann equation describes the evolution of the entire regime, encompassing both the system and its environment.
\begin{equation}
\frac{{d\rho }^{\prime }{\left( t\right) }}{{dt}}=-i\left[ {{H_{I}}\left(
t\right) ,\rho }^{\prime }{\left( t\right) }\right] .
\end{equation}
where ${\rho }^{\prime }{\left( t\right) }$ denotes the total density
matrix, and we can decompose the interaction Hamiltonian ${H_I} = \lambda \sum\limits_{j,\mu } {{V_j}\left( {{\omega _\mu }} \right)} \otimes B_j^x.$
With the standard procedure of Born-Markov approximation, we
can get the evolution of the reduced density matrix ${\rho \left( t\right) }$ as 
\begin{align}
\frac{{d\rho \left( t\right) }}{{dt}}=\sum\limits_{\mu ,i,j}{{\Gamma_{i,j}}\left( {\omega _{\mu }}\right) \left[ {{V_{j}(\omega_\mu)}\rho \left( t\right),V_{i}^{\dag }(\omega_\mu)}\right] }+\sum\limits_{\mu ,i,j}{\Gamma_{j,i}^{\ast }\left( {\omega _{\mu }}\right)\left[ {{V_{j}(\omega_\mu)},\rho \left( t\right) V_{i}^{\dag }(\omega_\mu)}\right] },
\label{A2}
\end{align}
where 
\begin{equation}
{\Gamma_{i,j}}\left( {\omega _{\mu }}\right) =\int\limits_{0}^{\infty }{ds{e^{i{%
\omega _{\mu }}s}}\left\langle {B_{i}^{x}\left( s\right) B_{j}^{x}}%
\right\rangle }
\end{equation}
is the forward Fourier transform of the reservoir correlation function $%
\left\langle {B_{i}^{x}\left( s\right) B_{j}^{x}}\right\rangle $. Actually
only the terms $\left\langle {B_{j}^{x}\left( s\right) B_{j}^{x}}\right\rangle $ are not zero, which can be expressed as
\begin{align}
Tr\left( {B_{j}^{x}\left( s\right) B_{j}^{x}{\rho _{B}}}\right)=\sum\limits_{n}{{{\left\vert \kappa_{j,n}\right\vert }^{2}}\left( {{e^{-i{\omega _{n}}s}}\left( {{{\bar{n}}_{j}}\left( {\omega _{n}}\right) +1}\right)
+{e^{i{\omega _{n}}s}}{{\bar{n}}_{j}}\left( {\omega _{n}}\right) }\right) }\hfill  \notag \\
=\frac{1}{\pi }\int\limits_{0}^{\infty }{{J_{j}}\left( \omega \right)
\left( {{e^{-i{\omega }s}}\left( {{{\bar{n}}_{j}}\left( {\omega }\right) +1}\right) +{e^{i{\omega }s}}{{\bar{n}}_{j}}\left( {\omega }\right) }\right)
d\omega },\hfill
\end{align}
so we can obtain that 
\begin{align}
{\Gamma_{j,j}}\left( {\omega _{\mu }}\right) \equiv {\Gamma _j}\left( {{\omega _\mu }} \right) = \int\limits_{0}^{\infty }{ds{e^{i{%
\omega _{\mu }}s}}Tr\left( {B_{j}^{x}\left( s\right) B_{j}^{x}{\rho _{B}}}%
\right) }\hfill  \notag \\
=\frac{1}{\pi }\int\limits_{0}^{\infty }{\left( {\int\limits_{0}^{\infty }{%
{e^{-i\left( {\omega -{\omega _{\mu }}}\right) s}}ds}}\right) {J_{j}}\left(
\omega \right) \left( {{{\bar{n}}_{j}}\left( \omega \right) +1}\right) }%
d\omega \notag \\
+\frac{1}{\pi }\int\limits_{0}^{\infty }{\left( {\int\limits_{0}^{\infty }{%
{e^{-i\left( {\ -{\omega _{\mu }}-\omega }\right) s}}ds}}\right) {J_{j}}%
\left( \omega \right) {{\bar{n}}_{j}}\left( \omega \right) }d\omega .\hfill
\end{align}
Using the Kramers-Kronig relations 
\begin{equation}
\int\limits_{0}^{\infty }{{e^{-i\left( {\omega -{\omega _{0}}}\right) s}}ds}=\pi \delta \left( {\omega -{\omega _{0}}}\right) -iP.V.\frac{{1}}{{\omega -{\omega _{0}}}},
\end{equation}
and by the properties of the delta function
\begin{align}
\int\limits_0^\infty  {\delta \left( {\omega  - {\omega _\mu }} \right){J_j}\left( \omega  \right)\left( {{{\overline n }_j}\left( \omega  \right) + 1} \right)d\omega }  = {J_j}\left( {{\omega _\mu }} \right)\left( {{{\overline n }_j}\left( {{\omega _\mu }} \right) + 1} \right),
\end{align}
\begin{align}
\int\limits_0^\infty  {\delta \left( {\omega  + {\omega _\mu }} \right){J_j}\left( \omega  \right){{\overline n }_j}\left( \omega  \right)d\omega }  = 0,
\end{align}
we can obtain
\begin{align}
{\Gamma _j}\left( {{\omega _\mu }} \right) ={J_{j}}\left( {\omega _{\mu }}%
\right) \left( {{\bar{n}}_{j}\left( {\omega _{\mu }}\right) +1}\right)+\frac{i}{\pi }P.V.\int_{0}^{\infty }{{J_{j}}\left( \omega \right) \left( {%
\frac{{{{\bar{n}}_{j}}\left( \omega \right) +1}}{{{\omega _{\mu }}-\omega }}+%
\frac{{{{\bar{n}}_{j}}\left( \omega \right) }}{{{\omega _{\mu }}+\omega }}}%
\right) d\omega,} \label{A9}
\end{align}
\begin{align}
{\Gamma _j}\left( {{-\omega _\mu }} \right) ={J_{j}}\left( {\omega _{\mu }%
}\right) \bar{n}\left( {\omega _{\mu }}\right) -\frac{i}{\pi }P.V.\int_{0}^{\infty }{{J_{j}}\left( \omega \right) \left( {%
\frac{{{{\bar{n}}_{j}}\left( \omega \right) }}{{{\omega _{\mu }}-\omega }}+%
\frac{{{{\bar{n}}_{j}}\left( \omega \right) +1}}{{{\omega _{\mu }}+\omega }}}%
\right) d\omega ,} \label{A10}
\end{align}
where the $P.V.$ denotes the Cauchy principal value. 

Reorganising the terms in Eq. (\ref{A2}) in Schrödinger picture, we eventually arrive at
\begin{align}
\frac{{d\rho }}{{dt}} =  - i\left[ {{H_S} + {H_{LS}},\rho } \right] + \sum\limits_j {{\mathcal{L}_j}\left( \rho  \right)} ,
\label{A11}
\end{align}
where
\begin{align}
{H_{LS}} = \sum\limits_j {\sum\limits_{\omega ,\omega '} {{\xi _j}\left( {\omega ,\omega '} \right)V_j^\dag \left( {\omega '} \right){V_j}\left( \omega  \right)} } ,
\end{align}
\begin{align}
{\mathcal{L}_j}\left( \rho  \right) = \sum\limits_{\omega ,\omega '} {{\gamma _j}\left( {\omega ,\omega '} \right)\left( {{V_j}\left( {{\omega }} \right)\rho V_j^\dag \left( {\omega '} \right) - \frac{1}{2}\left\{ {V_j^\dag \left( {\omega '} \right){V_j}\left( \omega  \right),\rho } \right\}} \right)}. 
\end{align}
If $\left| {\omega  - \omega '} \right| \gg \tau _R^{ - 1} \sim {\lambda ^2}$, we can apply the secular approximation here, ignoring the terms where $\omega$ and $\omega'$ are different, we get the Lindblad master equation as
\begin{align}
H_{LS}^{L} = \sum\limits_{j,\omega } {{\xi _j}\left( {\omega ,\omega } \right)V_j^\dag \left( \omega  \right){V_j}\left( \omega  \right)} ,
\label{A14}
\end{align}
\begin{align}
\mathcal{L}_j^{L}\left( \rho  \right) = \sum\limits_{\omega } {{\gamma _j}\left( {\omega ,\omega } \right)\left( {{V_j}\left( \omega  \right)\rho V_j^\dag \left( \omega  \right) - \frac{1}{2}\left\{ {V_j^\dag \left( \omega  \right){V_j}\left( \omega  \right),\rho } \right\}} \right)} .
\label{A15}
\end{align}
According to Eqs. (\ref{A9},\ref{A10}), we have
\begin{align}
{\gamma _j}\left( {\omega ,\omega } \right) = 2{G_j}\left( \omega  \right) = 2{J_j}\left( \omega  \right)\left( {{{\overline n }_j}\left( \omega  \right) + 1} \right),
\end{align}
\begin{align}
{\xi _j}\left( {\omega ,\omega } \right) = {S_j}\left( \omega  \right) = \frac{1}{\pi }P.V.\int_0^\infty  {{J_j}\left( {\omega '} \right)\left( {\frac{{{{\overline n }_j}\left( {\omega '} \right) + 1}}{{\omega  - \omega '}} + \frac{{{{\overline n }_j}\left( {\omega '} \right)}}{{\omega  + \omega '}}} \right)d\omega '} ,
\end{align}
where the KMS condition is satisfied, i.e., 
\begin{equation}
{\gamma _j}\left( { - \omega , - \omega } \right) = 2{G_j}\left( { - \omega } \right) = 2{e^{ - {\beta _j}\omega }}{G_j}\left( \omega  \right) = {e^{ - {\beta _j}\omega }}{\gamma _j}\left( {\omega ,\omega } \right).
\end{equation}

\section{The steady states of the Bloch-Redfield master equation}
\label{appendixB}
By substituting the eigenoperators into the master equation Eq. (\ref{A11}) and making simplification, we find that the evolution of the density matrix of the system decouples into two parts:
\[{\rho _S} = \left( {\begin{array}{*{20}{c}}
  {{\rho _{11}}}&{}&{}&{{\rho _{14}}} \\ 
  {}&{{\rho _{22}}}&{{\rho _{23}}}&{} \\ 
  {}&{{\rho _{32}}}&{{\rho _{33}}}&{} \\ 
  {{\rho _{41}}}&{}&{}&{{\rho _{44}}} 
\end{array}} \right),{\rho _O} = \left( {\begin{array}{*{20}{c}}
  {}&{{\rho _{12}}}&{{\rho _{13}}}&{} \\ 
  {{\rho _{21}}}&{}&{}&{{\rho _{24}}} \\ 
  {{\rho _{31}}}&{}&{}&{{\rho _{34}}} \\ 
  {}&{{\rho _{42}}}&{{\rho _{43}}}&{} 
\end{array}} \right),\]
$\rho _S$ consists of the elements on the primary and secondary diagonals of the density matrix, and $\rho _O$ consists of the remaining elements. They both evolved independently  as
\begin{align}
\frac{d}{{dt}}{\rho _S} = {M_{11}}{\rho _{11}} + {M_{22}}{\rho _{22}} + {M_{33}}{\rho _{33}} + {M_{44}}{\rho _{44}} + {M_{14}}{\rho _{14}} + {M_{23}}{\rho _{23}} + {M_{32}}{\rho _{32}} + {M_{41}}{\rho _{41}},
\label{B1}
\end{align}
\begin{align}
\frac{d}{{dt}}{\rho _O} = {M_{12}}{\rho _{12}} + {M_{13}}{\rho _{13}} + {M_{21}}{\rho _{21}} + {M_{24}}{\rho _{24}} + {M_{31}}{\rho _{31}} + {M_{34}}{\rho _{34}} + {M_{42}}{\rho _{42}} + {M_{43}}{\rho _{43}},
\label{B2}
\end{align}
where the explicit expressions for $M_{ij}$ are given as follows. First we introduce some notation to simplify the result. For $F \in \left\{ {G, S} \right\}$,
\[\begin{gathered}
  {\overline F _1}\left( \omega  \right) \equiv {F_1}\left( \omega  \right)\sin {\phi _ + }\cos {\phi _ + },\forall \omega ;{\overline F _2}\left( \omega  \right) \equiv {F_2}\left( \omega  \right)\sin {\phi _ - }\cos {\phi _ - },\forall \omega ; \hfill \\
  {\widetilde F_1}\left( \omega  \right) \equiv {F_1}\left( \omega  \right){\sin ^2}{\phi _ + },\omega  =  \pm {\omega _1};{\widetilde F_1}\left( \omega  \right) \equiv {F_1}\left( \omega  \right){\cos ^2}{\phi _ + },\omega  =  \pm {\omega _2}; \hfill \\
  {\widetilde F_2}\left( \omega  \right) \equiv {F_2}\left( \omega  \right){\cos ^2}{\phi _ - },\omega  =  \pm {\omega _1};{\widetilde F_2}\left( \omega  \right) \equiv {F_2}\left( \omega  \right){\sin ^2}{\phi _ - },\omega  =  \pm {\omega _2}; \hfill \\
  {\overline F _ \pm } = {\overline F _1} \pm {\overline F _2},{\widetilde F_ \pm } = {\widetilde F_1} \pm {\widetilde F_2}. \hfill \\ 
\end{gathered} \]
Then the expressions for $M_{ij}$ are
\[\begin{gathered}
  {M_{11}} = \left( {\begin{array}{*{20}{c}}
  { - 2{{\widetilde G}_ + }\left( {{\omega _1}} \right) - 2{{\widetilde G}_ + }\left( {{\omega _2}} \right)}&0&0&{{M_{11}}\left( {1,4} \right)} \\ 
  0&{2{{\widetilde G}_ + }\left( {{\omega _1}} \right)}&{{M_{11}}\left( {2,3} \right)}&0 \\ 
  0&{M_{11}^*\left( {2,3} \right)}&{2{{\widetilde G}_ + }\left( {{\omega _2}} \right)}&0 \\ 
  {M_{11}^*\left( {1,4} \right)}&0&0&0 
\end{array}} \right), \hfill \\
  {M_{11}}\left( {1,4} \right) = {\overline G _ + }\left( {{\omega _2}} \right) - {\overline G _ + }\left( {{\omega _1}} \right) + i{\overline S _ + }\left( {{\omega _1}} \right) - i{\overline S _ + }\left( {{\omega _2}} \right), \hfill \\
  {M_{11}}\left( {2,3} \right) = {\overline G _ - }\left( {{\omega _1}} \right) + {\overline G _ - }\left( {{\omega _2}} \right) + i{\overline S _ - }\left( {{\omega _1}} \right) - i{\overline S _ - }\left( {{\omega _2}} \right); \hfill \\ 
\end{gathered} \]

\[\begin{gathered}
  {M_{22}} = \left( {\begin{array}{*{20}{c}}
  {2{{\widetilde G}_ + }\left( { - {\omega _1}} \right)}&0&0&{{M_{22}}\left( {1,4} \right)} \\ 
  0&{ - 2{{\widetilde G}_ + }\left( { - {\omega _1}} \right) - 2{{\widetilde G}_ + }\left( {{\omega _2}} \right)}&{{M_{22}}\left( {2,3} \right)}&0 \\ 
  0&{M_{22}^*\left( {2,3} \right)}&0&0 \\ 
  {M_{22}^*\left( {1,4} \right)}&0&0&{2{{\widetilde G}_ + }\left( {{\omega _2}} \right)} 
\end{array}} \right), \hfill \\
  {M_{22}}\left( {1,4} \right) = {\overline G _ + }\left( { - {\omega _1}} \right) + {\overline G _ + }\left( {{\omega _2}} \right) + i{\overline S _ + }\left( { - {\omega _1}} \right) - i{\overline S _ + }\left( {{\omega _2}} \right), \hfill \\
  {M_{22}}\left( {2,3} \right) = {\overline G _ - }\left( {{\omega _2}} \right) - {\overline G _ - }\left( { - {\omega _1}} \right) + i{\overline S _ - }\left( { - {\omega _1}} \right) - i{\overline S _ - }\left( {{\omega _2}} \right); \hfill \\ 
\end{gathered} \]

\[\begin{gathered}
  {M_{33}} = \left( {\begin{array}{*{20}{c}}
  {2{{\widetilde G}_ + }\left( { - {\omega _2}} \right)}&0&0&{{M_{33}}\left( {1,4} \right)} \\ 
  0&0&{{M_{33}}\left( {2,3} \right)}&0 \\ 
  0&{M_{33}^*\left( {2,3} \right)}&{ - 2{{\widetilde G}_ + }\left( {{\omega _1}} \right) - 2{{\widetilde G}_ + }\left( { - {\omega _2}} \right)}&0 \\ 
  {M_{33}^*\left( {1,4} \right)}&0&0&{2{{\widetilde G}_ + }\left( {{\omega _1}} \right)} 
\end{array}} \right), \hfill \\
  {M_{33}}\left( {1,4} \right) =  - {\overline G _ + }\left( {{\omega _1}} \right) - {\overline G _ + }\left( { - {\omega _2}} \right) + i{\overline S _ + }\left( {{\omega _1}} \right) - i{\overline S _ + }\left( { - {\omega _2}} \right), \hfill \\
  {M_{33}}\left( {2,3} \right) = {\overline G _ - }\left( {{\omega _1}} \right) - {\overline G _ - }\left( { - {\omega _2}} \right) + i{\overline S _ - }\left( {{\omega _1}} \right) - i{\overline S _ - }\left( { - {\omega _2}} \right); \hfill \\ 
\end{gathered} \]

\[\begin{gathered}
  {M_{44}} = \left( {\begin{array}{*{20}{c}}
  0&0&0&{{M_{44}}\left( {1,4} \right)} \\ 
  0&{2{{\widetilde G}_ + }\left( { - {\omega _2}} \right)}&{{M_{44}}\left( {2,3} \right)}&0 \\ 
  0&{M_{44}^*\left( {2,3} \right)}&{2{{\widetilde G}_ + }\left( { - {\omega _1}} \right)}&0 \\ 
  {M_{44}^*\left( {1,4} \right)}&0&0&{ - 2{{\widetilde G}_ + }\left( { - {\omega _1}} \right) - 2{{\widetilde G}_ + }\left( { - {\omega _2}} \right)} 
\end{array}} \right), \hfill \\
  {M_{44}}\left( {1,4} \right) = {\overline G _ + }\left( { - {\omega _1}} \right) - {\overline G _ + }\left( { - {\omega _2}} \right) + i{\overline S _ + }\left( { - {\omega _1}} \right) - i{\overline S _ + }\left( { - {\omega _2}} \right), \hfill \\
  {M_{44}}\left( {2,3} \right) =  - {\overline G _ - }\left( { - {\omega _1}} \right) - {\overline G _ - }\left( { - {\omega _2}} \right) + i{\overline S _ - }\left( { - {\omega _1}} \right) - i{\overline S _ - }\left( { - {\omega _2}} \right); \hfill \\ 
\end{gathered} \]

\[\begin{gathered}
  {M_{14}} = \left( {\begin{array}{*{20}{c}}
  {M_{44}^*\left( {1,4} \right)}&0&0&{{M_{14}}\left( {1,4} \right)} \\ 
  0&{ - M_{33}^*\left( {1,4} \right)}&{{M_{14}}\left( {2,3} \right)}&0 \\ 
  0&{{M_{14}}\left( {3,2} \right)}&{ - M_{22}^*\left( {1,4} \right)}&0 \\ 
  0&0&0&{M_{11}^*\left( {1,4} \right)} 
\end{array}} \right), \hfill \\
  {M_{14}}\left( {1,4} \right) =  - \left( {{{\widetilde G}_ + }\left( {{\omega _1}} \right) + {{\widetilde G}_ + }\left( {{\omega _2}} \right) + {{\widetilde G}_ + }\left( { - {\omega _1}} \right) + {{\widetilde G}_ + }\left( { - {\omega _2}} \right)} \right) \hfill \\
   + i{\widetilde S_ + }\left( { - {\omega _1}} \right) + i{\widetilde S_ + }\left( { - {\omega _2}} \right) - i{\widetilde S_ + }\left( {{\omega _1}} \right) - i{\widetilde S_ + }\left( {{\omega _2}} \right) - 2i\beta , \hfill \\
  {M_{14}}\left( {2,3} \right) =  - \left( {{{\widetilde G}_ - }\left( {{\omega _1}} \right) + {{\widetilde G}_ - }\left( { - {\omega _1}} \right)} \right) + i{\widetilde S_ - }\left( { - {\omega _1}} \right) - i{\widetilde S_ - }\left( {{\omega _1}} \right), \hfill \\
  {M_{14}}\left( {3,2} \right) = {\widetilde G_ - }\left( {{\omega _2}} \right) + {\widetilde G_ - }\left( { - {\omega _2}} \right) + i{\widetilde S_ - }\left( {{\omega _2}} \right) - i{\widetilde S_ - }\left( { - {\omega _2}} \right); \hfill \\ 
\end{gathered} \]

\[\begin{gathered}
  {M_{23}} = \left( {\begin{array}{*{20}{c}}
  { - M_{44}^*\left( {2,3} \right)}&0&0&{M_{14}^*\left( {2,3} \right)} \\ 
  0&{M_{33}^*\left( {2,3} \right)}&{{M_{23}}\left( {2,3} \right)}&0 \\ 
  0&0&{M_{22}^*\left( {2,3} \right)}&0 \\ 
  {{M_{14}}\left( {3,2} \right)}&0&0&{ - M_{11}^*\left( {2,3} \right)} 
\end{array}} \right), \hfill \\
  {M_{23}}\left( {2,3} \right) =  - \left( {{{\widetilde G}_ + }\left( {{\omega _1}} \right) + {{\widetilde G}_ + }\left( {{\omega _2}} \right) + {{\widetilde G}_ + }\left( { - {\omega _1}} \right) + {{\widetilde G}_ + }\left( { - {\omega _2}} \right)} \right) \hfill \\
   - i{\widetilde S_ + }\left( { - {\omega _1}} \right) + i{\widetilde S_ + }\left( { - {\omega _2}} \right) + i{\widetilde S_ + }\left( {{\omega _1}} \right) - i{\widetilde S_ + }\left( {{\omega _2}} \right) - 2i\alpha ; \hfill \\ 
\end{gathered} \]

\[{M_{41}} = M_{14}^\dag ,{M_{32}} = M_{23}^\dag ;\]

\[{M_{12}} = \left( {\begin{array}{*{20}{c}}
  0&{{M_{12}}\left( {1,2} \right)}&{{M_{22}}\left( {2,3} \right)}&0 \\ 
  {{M_{12}}\left( {2,1} \right)}&0&0&{{M_{12}}\left( {2,4} \right)} \\ 
  {{M_{12}}\left( {3,1} \right)}&0&0&{2{{\widetilde G}_ - }\left( {{\omega _2}} \right)} \\ 
  0&{M_{11}^*\left( {1,4} \right)}&0&0 
\end{array}} \right),\]
\[{M_{12}}\left( {1,2} \right) =  - \left( {{{\widetilde G}_ + }\left( { - {\omega _1}} \right) + {{\widetilde G}_ + }\left( {{\omega _1}} \right) + 2{{\widetilde G}_ + }\left( {{\omega _2}} \right)} \right) + i\left( {\alpha  - \beta  + {{\widetilde S}_ + }\left( { - {\omega _1}} \right) - {{\widetilde S}_ + }\left( {{\omega _1}} \right)} \right),\]
\[\begin{gathered}
  {M_{12}}\left( {2,1} \right) = {\widetilde G_ + }\left( {{\omega _1}} \right) + {\widetilde G_ + }\left( { - {\omega _1}} \right) + i{\widetilde S_ + }\left( {{\omega _1}} \right) - i{\widetilde S_ + }\left( { - {\omega _1}} \right), \hfill \\
  {M_{12}}\left( {2,4} \right) = {\overline G _ + }\left( {{\omega _1}} \right) + {\overline G _ + }\left( {{\omega _2}} \right) + i{\overline S _ + }\left( {{\omega _1}} \right) - i{\overline S _ + }\left( {{\omega _2}} \right), \hfill \\
  {M_{12}}\left( {3,1} \right) = {\overline G _ - }\left( { - {\omega _1}} \right) + {\overline G _ - }\left( {{\omega _2}} \right) + i{\overline S _ - }\left( {{\omega _2}} \right) - i{\overline S _ - }\left( { - {\omega _1}} \right); \hfill \\ 
\end{gathered} \]

\[{M_{13}} = \left( {\begin{array}{*{20}{c}}
  0&{M_{33}^*\left( {2,3} \right)}&{{M_{13}}\left( {1,3} \right)}&0 \\ 
  {{M_{13}}\left( {2,1} \right)}&0&0&{ - 2{{\widetilde G}_ - }\left( {{\omega _1}} \right)} \\ 
  {{M_{13}}\left( {3,1} \right)}&0&0&{{M_{13}}\left( {3,4} \right)} \\ 
  0&0&{M_{11}^*\left( {1,4} \right)}&0 
\end{array}} \right),\]
\[{M_{13}}\left( {1,3} \right) =  - \left( {{{\widetilde G}_ + }\left( {{\omega _2}} \right) + {{\widetilde G}_ + }\left( { - {\omega _2}} \right) + 2{{\widetilde G}_ + }\left( {{\omega _1}} \right)} \right) - i\left( {\alpha  + \beta  + {{\widetilde S}_ + }\left( {{\omega _2}} \right) - {{\widetilde S}_ + }\left( { - {\omega _2}} \right)} \right),\]
\[\begin{gathered}
  {M_{13}}\left( {2,1} \right) = {\overline G _ - }\left( {{\omega _1}} \right) + {\overline G _ - }\left( { - {\omega _2}} \right) + i{\overline S _ - }\left( {{\omega _1}} \right) - i{\overline S _ - }\left( { - {\omega _2}} \right), \hfill \\
  {M_{13}}\left( {3,1} \right) = {\widetilde G_ + }\left( {{\omega _2}} \right) + {\widetilde G_ + }\left( { - {\omega _2}} \right) + i{\widetilde S_ + }\left( {{\omega _2}} \right) - i{\widetilde S_ + }\left( { - {\omega _2}} \right), \hfill \\
  {M_{13}}\left( {3,4} \right) =  - {\overline G _ + }\left( {{\omega _1}} \right) - {\overline G _ + }\left( {{\omega _2}} \right) + i{\overline S _ + }\left( {{\omega _1}} \right) - i{\overline S _ + }\left( {{\omega _2}} \right); \hfill \\ 
\end{gathered} \]

\[{M_{24}} = \left( {\begin{array}{*{20}{c}}
  0&{{M_{24}}\left( {1,2} \right)}&{ - 2{{\widetilde G}_ - }\left( { - {\omega _1}} \right)}&0 \\ 
  {M_{44}^*\left( {1,4} \right)}&0&0&{{M_{24}}\left( {2,4} \right)} \\ 
  0&0&0&{M_{22}^*\left( {2,3} \right)} \\ 
  0&{{M_{13}}\left( {3,1} \right)}&{{M_{24}}\left( {4,3} \right)}&0 
\end{array}} \right),\]
\[{M_{24}}\left( {2,4} \right) =  - \left( {{{\widetilde G}_ + }\left( {{\omega _2}} \right) + {{\widetilde G}_ + }\left( { - {\omega _2}} \right) + 2{{\widetilde G}_ + }\left( { - {\omega _1}} \right)} \right) - i\left( {\alpha  + \beta  + {{\widetilde S}_ + }\left( {{\omega _2}} \right) - {{\widetilde S}_ + }\left( { - {\omega _2}} \right)} \right),\]
\[\begin{gathered}
  {M_{24}}\left( {1,2} \right) = {\overline G _ + }\left( { - {\omega _1}} \right) + {\overline G _ + }\left( { - {\omega _2}} \right) + i{\overline S _ + }\left( { - {\omega _1}} \right) - i{\overline S _ + }\left( { - {\omega _2}} \right), \hfill \\
  {M_{24}}\left( {4,3} \right) =  - {\overline G _ - }\left( { - {\omega _1}} \right) - {\overline G _ - }\left( {{\omega _2}} \right) + i{\overline S _ - }\left( { - {\omega _1}} \right) - i{\overline S _ - }\left( {{\omega _2}} \right); \hfill \\ 
\end{gathered} \]

\[{M_{34}} = \left( {\begin{array}{*{20}{c}}
  0&{2{{\widetilde G}_ - }\left( { - {\omega _2}} \right)}&{{M_{34}}\left( {1,3} \right)}&0 \\ 
  0&0&0&{{M_{33}}\left( {2,3} \right)} \\ 
  {M_{44}^*\left( {1,4} \right)}&0&0&{{M_{34}}\left( {3,4} \right)} \\ 
  0&{{M_{34}}\left( {4,2} \right)}&{{M_{12}}\left( {2,1} \right)}&0 
\end{array}} \right),\]
\[{M_{34}}\left( {3,4} \right) =  - \left( {{{\widetilde G}_ + }\left( {{\omega _1}} \right) + {{\widetilde G}_ + }\left( { - {\omega _1}} \right) + 2{{\widetilde G}_ + }\left( { - {\omega _2}} \right)} \right) + i\left( {\alpha  - \beta  + {{\widetilde S}_ + }\left( { - {\omega _1}} \right) - {{\widetilde S}_ + }\left( {{\omega _1}} \right)} \right),\]
\[\begin{gathered}
  {M_{34}}\left( {1,3} \right) =  - {\overline G _ + }\left( { - {\omega _1}} \right) - {\overline G _ + }\left( { - {\omega _2}} \right) + i{\overline S _ + }\left( { - {\omega _1}} \right) - i{\overline S _ + }\left( { - {\omega _2}} \right), \hfill \\
  {M_{34}}\left( {4,2} \right) =  - {\overline G _ - }\left( {{\omega _1}} \right) - {\overline G _ - }\left( { - {\omega _2}} \right) + i{\overline S _ - }\left( { - {\omega _2}} \right) - i{\overline S _ - }\left( {{\omega _1}} \right); \hfill \\ 
\end{gathered} \]

\[{M_{21}} = M_{12}^\dag ,{M_{31}} = M_{13}^\dag ,{M_{42}} = M_{24}^\dag ,{M_{43}} = M_{34}^\dag .\]
Notice that all the $M_{ij}$ here are traceless,  because $\frac{d}{{dt}}Tr {{\rho _S}}  = \frac{d}{{dt}}Tr {{\rho _O}}  = 0$. To obtain the steady-state solution, we set $\frac{{d\rho }}{{dt}} = 0$, that is $\frac{{d{\rho _S}}}{{dt}} = 0$ and $\frac{{d{\rho _O}}}{{dt}} = 0$. For the former, we can calculate ${\rho _S}$ in a perturbative way:
\begin{equation}
\rho _S^{steady} = \rho _{pd}^{\left( 0 \right)} + \rho _{sd}^{\left( 0 \right)} + {\lambda ^2}\rho _{pd}^{\left( 2 \right)} + {\lambda ^2}\rho _{sd}^{\left( 2 \right)} + O\left( {{\lambda ^4}} \right),
\end{equation}
where ${\rho _{pd}}$ stands for the primary diagonal elements, ${\rho _{sd}}$ stands for the secondary diagonal elements, and we know that $Tr\rho_{pd}^{(0)}=1, Tr\rho_{pd}^{(2)}=0$. In combination with Eq. (\ref{B1}), given that both $G(\omega)$ and $S(\omega)$ are of order ${\lambda ^2}$, the coefficient of ${\lambda ^0}$ in the equation gives
\[2i\left( {\begin{array}{*{20}{c}}
  {}&{}&{}&{ - \beta \rho _{14}^{\left( 0 \right)}} \\ 
  {}&{}&{ - \alpha \rho _{23}^{\left( 0 \right)}}&{} \\ 
  {}&{\alpha \rho _{32}^{\left( 0 \right)}}&{}&{} \\ 
  {\beta \rho _{41}^{\left( 0 \right)}}&{}&{}&{} 
\end{array}} \right) = 0.\]
For the non-degenerate cases (the degenerate case is discussed in Appendix \ref{appendixC}), $\alpha \ne 0,$ this means $\rho _{sd}^{\left( 0 \right)} = 0$; the coefficient of ${\lambda ^2}$ in the diagonal gives
\[\begin{gathered}
   - \left( {{{\widetilde G}_ + }\left( {{\omega _1}} \right) + {{\widetilde G}_ + }\left( {{\omega _2}} \right)} \right)\rho _{11}^{\left( 0 \right)} + {\widetilde G_ + }\left( { - {\omega _1}} \right)\rho _{22}^{\left( 0 \right)} + {\widetilde G_ + }\left( { - {\omega _2}} \right)\rho _{33}^{\left( 0 \right)} = 0, \hfill \\
  {\widetilde G_ + }\left( {{\omega _1}} \right)\rho _{11}^{\left( 0 \right)} - \left( {{{\widetilde G}_ + }\left( { - {\omega _1}} \right) + {{\widetilde G}_ + }\left( {{\omega _2}} \right)} \right)\rho _{22}^{\left( 0 \right)} + {\widetilde G_ + }\left( { - {\omega _2}} \right)\rho _{44}^{\left( 0 \right)} = 0, \hfill \\
  {\widetilde G_ + }\left( {{\omega _2}} \right)\rho _{11}^{\left( 0 \right)} - \left( {{{\widetilde G}_ + }\left( {{\omega _1}} \right) + {{\widetilde G}_ + }\left( { - {\omega _2}} \right)} \right)\rho _{33}^{\left( 0 \right)} + {\widetilde G_ + }\left( { - {\omega _1}} \right)\rho _{44}^{\left( 0 \right)} = 0, \hfill \\
  {\widetilde G_ + }\left( {{\omega _2}} \right)\rho _{22}^{\left( 0 \right)} + {\widetilde G_ + }\left( {{\omega _1}} \right)\rho _{33}^{\left( 0 \right)} - \left( {{{\widetilde G}_ + }\left( { - {\omega _1}} \right) + {{\widetilde G}_ + }\left( { - {\omega _2}} \right)} \right)\rho _{44}^{\left( 0 \right)} = 0. \hfill \\ 
\end{gathered} \]
One can find that only three of these four equations are independent of each other. Considering that $Tr\rho _{pd}^{\left( 0 \right)} = 1,$ we obtain
\[
\rho_{pd}^{(0)} = \frac{1}{N}
\begin{pmatrix}
\widetilde{G}_+(-\omega_1)\widetilde{G}_+(-\omega_2) & & & \\
& \widetilde{G}_+(\omega_1)\widetilde{G}_+(-\omega_2) & & \\
& & \widetilde{G}_+(-\omega_1)\widetilde{G}_+(\omega_2) & \\
& & & \widetilde{G}_+(\omega_1)\widetilde{G}_+(\omega_2)
\end{pmatrix},
\]
with $N = \left( {{{\widetilde G}_ + }\left( {{\omega _1}} \right) + {{\widetilde G}_ + }\left( { - {\omega _1}} \right)} \right)\left( {{{\widetilde G}_ + }\left( {{\omega _2}} \right) + {{\widetilde G}_ + }\left( { - {\omega _2}} \right)} \right),$ which happens to coincide with the steady-state solution of the Lindblad  master equation (\ref{A14})(\ref{A15}). The coefficient of ${\lambda ^2}$ in the secondary diagonal elements gives ${\lambda ^2}\rho _{sd}^{\left( 2 \right)}$ 
\begin{gather}
  {\lambda ^2}\rho _{14}^{\left( 2 \right)} = {\lambda ^2}\rho _{41}^{\left( 2 \right)*} = \frac{1}{{2i\beta }}\sum\limits_{i = 1}^4 {{M_{ii}}\left( {1,4} \right)\rho _{ii}^{\left( 0 \right)}} , \label{B4} \hfill \\
  {\lambda ^2}\rho _{23}^{\left( 2 \right)} = {\lambda ^2}\rho _{32}^{\left( 2 \right)*} = \frac{1}{{2i\alpha }}\sum\limits_{i = 1}^4 {{M_{ii}}\left( {2,3} \right)\rho _{ii}^{\left( 0 \right)}} . \label{B5} \hfill
\end{gather}
Similarly, we can obtain four equations about ${\lambda ^2}\rho _{pd}^{\left( 2 \right)}$ by the coefficient of ${\lambda ^4}$ in the primary diagonal elements, but only three of them are independent of each other, and considering $Tr\rho _{pd}^{\left( 2 \right)} = 0$ we can then compute ${\lambda ^2}\rho _{pd}^{\left( 2 \right)}$.

As for $\frac{{d{\rho _O}}}{{dt}} = 0$, we find that if ${{\rho _O}}$ is written as a column vector $\left| {{\rho _O}} \right\rangle $, then Eq. (\ref{B2}) corresponds to the equation $\frac{{d\left| {{\rho _O}} \right\rangle }}{{dt}} = \mathcal{M}\left| {{\rho _O}} \right\rangle $ in which the eight-dimensional matrix $\mathcal{M}$ is a diagonally dominant matrix: 
\[\left| {{\mathcal{M}_{kk}}} \right| \sim \beta  \gg {\lambda ^2} \sim \sum\limits_{l \ne k} {\left| {{\mathcal{M}_{kl}}} \right|}, \]
so the determinant of $\mathcal{M}$ is not zero, i.e. $\mathcal{M}$ is invertible, thus the solution to $\mathcal{M}\left| {{\rho _O}} \right\rangle  = 0$ must be a zero vector.

\section{The heat current for the symmetric case as $g\rightarrow 0$ }
\label{appendixC}
For the symmetric case, in the limit of $g\rightarrow 0$, we obtain
$\varphi  \to 0,\theta  \to \frac{\pi }{2},\alpha  \to 0,\beta  \to \varepsilon, {\omega _1} = {\omega _2} \to \varepsilon$. Substituting these into Eq. (\ref{lindcurrent}), we get 
\begin{equation}
{\mathcal{J}^{L}} = \frac{{J\left( \varepsilon  \right)\left( {{{\overline n }_1}\left( \varepsilon  \right) - {{\overline n }_2}\left( \varepsilon  \right)} \right)}}{{1 + {{\overline n }_1}\left( \varepsilon  \right) + {{\overline n }_2}\left( \varepsilon  \right)}}\left( {\varepsilon  + \delta } \right)
\end{equation}
with $\delta  = \left( {{S_1}\left( \varepsilon  \right) + {S_2}\left( \varepsilon  \right) - {S_1}\left( { - \varepsilon } \right) - {S_2}\left( { - \varepsilon } \right)} \right)/2$, where we have taken $\gamma_1=\gamma_2$ for simplicity, so that $J_1(\varepsilon)=J_2(\varepsilon) \equiv J(\varepsilon)$. $\mathcal{J}^{L}$ is a non-zero value at $T_1 \ne T_2$, which is unphysical. In the limit $g \to 0$, the system decouples into two independent subsystems, effectively eliminating any interaction between the thermal reservoirs. Consequently, the heat current must vanish to maintain consistency with thermodynamic principles. 

But in the BR master equation, substituting these into $M_{ij}$, the exact steady-state solution can be achieved by the non-perturbative direct solution method as
\[\rho_S  = \frac{1}{{\left( {1 + 2{n_1}} \right)\left( {1 + 2{n_2}} \right)}}\left( {\begin{array}{*{20}{c}}
  {{n_1}{n_2}}&{}&{}&{} \\ 
  {}&{{n_1}{n_2} + \frac{{{n_1} + {n_2}}}{2}}&{\frac{{{n_2} - {n_1}}}{2}}&{} \\ 
  {}&{\frac{{{n_2} - {n_1}}}{2}}&{{n_1}{n_2} + \frac{{{n_1} + {n_2}}}{2}}&{} \\ 
  {}&{}&{}&{\left( {{n_1} + 1} \right)\left( {{n_2} + 1} \right)} 
\end{array}} \right),\]
where ${n_1} = {\overline n _1}\left( \varepsilon  \right),{n_2} = {\overline n _2}\left( \varepsilon  \right),$ it can be seen that this steady-state solution coincides with a unitary transformation of the direct product of the local thermal state of the two TLAs, which implies the steady-state solution of the local master equation. From Eq. (\ref{BRcurrent}), it can be calculated that the heat current is zero, which is in agreement with the previous discussion. This shows that in our model, the BR master equation degenerates into the local Lindblad master equation only in the symmetric case, and if the internal coupling of the system is very weak, for $\varepsilon_1 \ne \varepsilon_2$ the BR master equation does not degenerate into the local Lindblad master equation even though the internal coupling of the system is very weak.

This explains why the heat current predicted by the local Lindblad master equation at $\varepsilon_1 \ne \varepsilon_2$ sometimes violates the second law of thermodynamics \cite{Novotný_2002, Levy_2014}, since the BR master equation in this case does not approximate the local Lindblad master equation. This is consistent with the results obtained by a recently developed master equation based on subsystem eigenstates \cite{PhysRevE.107.014108}, which provides a second-order approximation to the BR master equation as $g$ tends to zero.
\end{widetext}
\end{document}